\documentclass{article}
\usepackage{arxiv}

\usepackage[utf8]{inputenc} 
\usepackage[T1]{fontenc} 
\usepackage{hyperref} 
\usepackage{url} 
\usepackage{booktabs} 
\usepackage{multirow} 
\usepackage{amsfonts} 
\usepackage{nicefrac} 
\usepackage{microtype} 
\usepackage{graphicx}
\usepackage[numbers,sort&compress]{natbib}
\usepackage{doi}
\usepackage{amsmath,amssymb}
\usepackage{bm}            
\newcommand{\tb}[1]{\textbf{\boldmath #1}}
\usepackage{caption} 
\usepackage{cleveref} 
\usepackage[ruled,vlined,linesnumbered]{algorithm2e} 
\SetKwInput{KwInput}{Input}
\SetKwInput{KwOutput}{Output}
\SetKwInput{KwHyper}{Hyperparameters}

\title{Global kilometre-scale tropical cyclone inner-core vector winds from sparse scalar CYGNSS observations}

\newif\ifuniqueAffiliation
\uniqueAffiliationtrue

\author{\begin{minipage}{0.92\textwidth}
	\centering
	\normalfont\linespread{1.2}\selectfont
	Xinhai Han$^{1,2,3}$, Xiaohui Li$^{2,*}$, Jingsong Yang$^{2,3,4,*}$, Zeyi Niu$^{1}$, Guoqi Han$^{5}$, Jiuke Wang$^{6}$, Wei Huang$^{1}$, Yunxia Zheng$^{1}$, Hanyue Ni$^{2,3}$, Yiqi Wang$^{2}$, Wei Tao$^{7}$, Lotfi Aouf$^{8}$, Shaoliang Peng$^{9,10}$, Dake Chen$^{2,3,4}$\\[8pt]
	Corresponding author: Xiaohui Li (lixiaohui1991@live.cn), Jingsong Yang (jsyang@sio.org.cn)\\[8pt]
	$^{1}$Shanghai Typhoon Institute, and Key Laboratory of Numerical Modeling for Tropical Cyclone of the China Meteorological Administration, Shanghai 200030, China\\
	$^{2}$State Key Laboratory of Satellite Ocean Environment Dynamics, Second Institute of Oceanography, Ministry of Natural Resources, Hangzhou 310012, China\\
	$^{3}$School of Oceanography, Shanghai Jiao Tong University, Shanghai 200030, China\\
	$^{4}$Southern Marine Science and Engineering Guangdong Laboratory (Zhuhai), Zhuhai 519082, China\\
	$^{5}$Fisheries and Oceans Canada, Institute of Ocean Sciences, Sidney, V8L 4B2, Canada\\
	$^{6}$School of Artificial Intelligence, Sun Yat-sen University, Zhuhai 519082, China\\
	$^{7}$School of Electronic Information and Electrical Engineering, Shanghai Jiao Tong University, Shanghai 200240, China\\
	$^{8}$Department of Marine and Oceanography, 31057 Toulouse, France\\
	$^{9}$College of Computer Science and Electronic Engineering, Hunan University, Changsha 410082, China\\
	$^{10}$State Key Laboratory of Chemo and Biosensing, Hunan University, Changsha 410082, China\\[8pt]
	$^{\dagger}$Affiliations 1 and 2 contributed equally to this work.
\end{minipage}}

\renewcommand{\headeright}{A Manuscript}
\renewcommand{\undertitle}{  }
\renewcommand{\shorttitle}{\textit{ }}

\hypersetup{
pdftitle={Global kilometre-scale tropical cyclone inner-core vector winds from sparse scalar CYGNSS observations},
pdfsubject={cs.AI, physics.ao-ph},
pdfauthor={Xinhai Han, Xiaohui Li, Jingsong Yang, Zeyi Niu, Guoqi Han, Jiuke Wang, Wei Huang, Yunxia Zheng, Hanyue Ni, Yiqi Wang, Wei Tao, Lotfi Aouf, Shaoliang Peng, Dake Chen},
pdfkeywords={Tropical Cyclone, CYGNSS, Diffusion Model, Data Assimilation, Wind Field Reconstruction},
colorlinks=true,
linkcolor=black,
citecolor=black,
urlcolor=black,
}

\begin{document}
\maketitle

\begin{abstract}
	Tropical cyclone (TC) inner-core surface wind vectors underpin intensity forecasting and storm-surge prediction, yet direct observations remain scarce: routine aircraft reconnaissance is confined to the North Atlantic and Eastern Pacific and, even there, samples each storm only episodically. CYGNSS is the only satellite that penetrates heavy precipitation to measure inner-core surface winds, but delivers directionless scalar wind speeds and is assimilated by no operational analysis system. Here we show that the full 10\,m vector wind field inside the TC inner core can be reconstructed globally at 1.5\,km resolution from sparse CYGNSS scalar observations alone, by generalising score-based diffusion assimilation to a nonlinear observation operator and injecting three TC boundary-layer constraints; we further propose a CYGNSS-intrinsic Observation Coverage Sufficiency (OCS) criterion that flags reliable reconstructions without external references. Applied to 4,955 snapshots of 249 TCs across all six active basins (2020--2022), the reconstructions reduce systematic $V_{\max}$ bias against IBTrACS best-track by $\sim$79\% and $\sim$75\% relative to ERA5 and CCMP. Independent Tail Doppler Radar validation (47 storms) yields a wind speed RMSE of 6.9\,m\,s$^{-1}$ on the 23 coverage-sufficient cases (7.5\,m\,s$^{-1}$ overall); ablation across the full sample shows that the physical constraints cut wind-direction RMSE by 60\% without degrading speed accuracy. The framework further supports joint assimilation of heterogeneous observations: adding only 11 dropsonde vectors to CYGNSS for TC FIONA (2022) reduces the cross-eye profile RMSE by 42\%, outlining a practical pathway for fusing CYGNSS with SFMR, SAR and scatterometer data. The result is a globally consistent, observation-anchored kilometre-scale description of TC inner-core vector winds across all six active basins, including those without routine aircraft reconnaissance.
\end{abstract}


\section{Introduction}
\label{sec:introduction}

Tropical cyclone (TC) sea surface 10\,m wind fields underpin operational forecasting and risk assessment, with structural parameters---maximum sustained wind speed ($V_{\max}$), Radius of Maximum Wind (RMW), and 34/50/64-kt wind radii (R34, R50, R64)---directly controlling storm-surge forecasting, ocean wave modeling, and engineering wind-load estimation \cite{knaff2007tcradii, demaria2005ships, vickery2009hurricane}, as well as downstream applications such as coastal water-level prediction \cite{han2025waterlevel, han2023olar}. In the TC boundary layer, the tangential and radial wind components jointly determine the inflow-angle structure that is closely tied to intensity change \cite{kepert2001tcbl, zhang2012inflow}, so characterizing the two-dimensional (2D) vector wind field---rather than scalar wind speed alone---is essential. High-resolution vector winds also constrain TC-ocean interaction studies: \citet{guan2026weak} used global drifting buoys to show that TC-induced inner-core sea-surface cooling is much weaker than satellite/model estimates, an analysis that hinges on access to such fields. However, current global reanalyses and satellite-merged products such as ERA5 \cite{hersbach2020era5} and the Cross-Calibrated Multi-Platform (CCMP) dataset \cite{atlas2011ccmp, mears2019ccmp} have only $\sim$0.25$^\circ$ ($\sim$25\,km) resolution: TC inner-core structures (eye, eyewall, inner rainbands) are severely smoothed and $V_{\max}$ is systematically underestimated. \citet{dulac2024era5tc} found that ERA5 peak TC intensity averages only $\sim$61\% of International Best Track Archive for Climate Stewardship (IBTrACS) values, with underestimation exceeding 40\% for Cat\,5 TCs; \citet{liu2025era5tc} reported ERA5 inner-core wind biases beyond $-11$\,m\,s$^{-1}$ versus SFMR; and CCMP underestimates TC winds by $\sim$6.5\% / 9\% relative to SMAP / SAR, requiring corrections up to 15\,m\,s$^{-1}$ in the inner core \cite{rong2024ccmphighwind}. This resolution and intensity gap motivates high-resolution TC wind-field reconstruction techniques.

Existing direct observation methods for TC inner-core wind fields each provide crucial local information but suffer from limited spatiotemporal coverage. Dropsondes are the gold standard in situ instrument, directly measuring 10\,m wind speed and direction at $\sim$1--2\,m\,s$^{-1}$ accuracy via on-board GPS \cite{franklin2003dropsonde, hock2006ncar}, but they rely on aircraft reconnaissance, currently routine only in the North Atlantic and Eastern Pacific \cite{aberson2006thirtyyears}, with the Western North Pacific, Northern Indian Ocean, and Southern Hemisphere essentially uncovered; even within reconnaissance basins, a single mission samples each storm only sparsely. Tail Doppler Radar (TDR) supplies $\sim$2\,km 3D inner-core composites \cite{marks1992norbert, reasor2013tdrmerge}, and the Stepped Frequency Microwave Radiometer (SFMR) provides $\sim$3--4\,m\,s$^{-1}$ along-track surface winds important for intensity calibration \cite{uhlhorn2007sfmr}, but both are tied to the same airborne platforms and their continuity is constrained by flight schedules and radar quality. From space, scatterometers (ASCAT \cite{figa2002ascat}, QuikSCAT \cite{hoffman2003quikscat}) are the only operational source of global sea-surface vector winds at $\sim$12.5--25\,km, but their Ku/C-band signals suffer severe rain attenuation in the eyewall \cite{stiles2014rain} and their effective resolution cannot resolve eye/eyewall fine structure. Synthetic Aperture Radar (SAR) provides sub-kilometre scalar wind-speed snapshots (native $\sim$500\,m, typically smoothed to $\sim$3\,km for 1-min sustained winds) \cite{mouche2017sar, li2013sar} sufficient to resolve eyewalls, but its polar-orbiting revisit is $\sim$12 days for a single satellite (a few days even with multi-constellations such as Sentinel-1 / RADARSAT) and current SAR products carry no wind-direction information. In short, airborne platforms are constrained by flight plans and regional deployments, while spaceborne scatterometers and SAR are restricted by orbital revisits and precipitation interference, leaving a complementary niche for high-revisit, globally deployable sensors.

Operational TC-specific models---the Hurricane Weather Research and Forecasting (HWRF) system and its successor the Hurricane Analysis and Forecast System (HAFS)---produce $\sim$1.5--2\,km $(u,v)$ analyses for every active TC at 6-hourly intervals and represent the state of the art in TC-specific guidance. However, their inner-core surface wind structures are shaped primarily by vortex initialisation (VI) procedures (synthetic bogus vortices or corrected previous-cycle forecasts) rather than by direct assimilation of sea-surface wind observations \cite{liu2020vortex}; TDR radial winds are assimilated in the inner core only when reconnaissance data exist \cite{lu2017impact}, applying to only $\sim$25\% of global TCs. For the majority of storms worldwide, HWRF/HAFS inner-core winds therefore reflect model physics and VI configuration rather than observationally constrained surface fields. Crucially, CYGNSS surface wind speeds---the only spaceborne observations that penetrate heavy precipitation to sample the TC inner core---are not assimilated by any current operational analysis system (HWRF/HAFS, ERA5, CCMP), leaving a significant observational resource systematically unexploited.

Launched in 2016, the Cyclone Global Navigation Satellite System (CYGNSS) constellation complements these gaps with high temporal revisits ($\sim$2.8\,h median) and global tropical coverage \cite{ruf2016cygnss, ruf2019cygnss}, exploiting GPS sea-surface reflections (GNSS Reflectometry) whose L-band signals penetrate heavy precipitation \cite{clarizia2014gnssr, morris2017cygnss} and remain valid where rain-attenuated scatterometer retrievals fail. The growing GNSS-R fleet---FY-3E GNOS-II \cite{huang2022fy3e, han2025fy3e, han2024igarss} and Tianmu-1 \cite{huang2025tianmu1}---together with recent gap-free \cite{du2026gapfree} and physics-constrained \cite{sun2026physics} reconstructions, is rapidly expanding this observational pillar. However, GNSS-R has three intrinsic limitations for TC vector-wind analysis. First, it measures only \emph{scalar} wind speed: the sea-surface scattering cross-section has weak sensitivity to wind direction, and \citet{pascual2021gnssrwinddir} showed that this sensitivity, marginally detectable at moderate wind speeds, decays sharply at high wind speeds in TC environments and falls far short of operational direction retrieval; consequently CYGNSS standard L2 products (both YSLF and Fully Developed Seas) carry scalar wind speed only, while vector winds---especially the inflow-angle structure---are essential for TC boundary-layer dynamics and intensity change \cite{zhang2012inflow, kepert2001tcbl}. Second, specular-point sampling is extremely sparse: a single pass yields only a handful of points within a TC, and the existing CYGNSS L3 Storm-Centric Gridded (SCG) product \cite{mayers2023scg}---which aggregates multi-pass specular points within a $\pm$6\,h window in TC-relative coordinates via inverse-variance weighted averaging on a $0.1^\circ$ ($\sim$11\,km) grid---is still scalar, track-banded with discontinuous coverage, and limited to the North Atlantic and Eastern Pacific at intensities $\geq$65\,kt. Third, even within the YSLF product specifically optimised for TC environments \cite{ruf2024characterization}, retrieval accuracy degrades sharply once wind speed exceeds $\sim$30\,m\,s$^{-1}$ as the L-band cross-section saturates, with errors growing from $\sim$2\,m\,s$^{-1}$ at low wind speeds to $\sim$8--13\,m\,s$^{-1}$ in extreme regimes \cite{ruf2019cygnss, ruf2024characterization}---precisely where inner-core observations matter most---so any reconstruction framework must adaptively down-weight unreliable retrievals rather than treat all observations as equally informative. The combined consequence is a severely ill-posed scalar-to-vector inverse problem: for any observed wind speed $y$, infinitely many $(u,v)$ satisfy $\sqrt{u^2+v^2}=y$, leaving the directional degrees of freedom unidentifiable. This dimensional leap from scalar to vector is the central scientific challenge in evolving CYGNSS from a wind-speed product into a vector wind-field product, and no existing method converts the CYGNSS record into spatially continuous, global, kilometre-scale TC \emph{vector} wind fields.

Existing reconstruction methods cannot overcome this fundamental challenge. Parametric vortex models such as the Holland (1980) profile \cite{holland1980vortex} impose axisymmetric assumptions and fixed inflow-angle parameterisations, failing to capture real-TC asymmetries, with inflow-angle errors exceeding 30$^\circ$ \cite{kepert2001tcbl}. Cubic-spline-type interpolation can grid scattered CYGNSS specular-point wind speeds, but---being driven only by scalar inputs---is structurally unable to recover $(u,v)$ components. Recent deep-learning approaches, including generative adversarial networks, have demonstrated TC wind-field reconstruction from SAR imagery \cite{han2023rs, li2024tgrs}, 3D wind-field reconstruction from sparse dropsonde profiles \cite{han2025grl}, multi-source 3D TC structure fusion \cite{eusebi2024tcrecon}, and score-based diffusion 3D reconstruction from sparse multi-variable observations \cite{han2026physics}; yet each inherits its observation source's coverage limits---aircraft inputs (dropsondes, TDR, in-situ profiles) carry direction information but are routinely available only over the North Atlantic and Eastern Pacific and sample each storm only episodically; spaceborne SAR is globally deployable but has multi-day per-TC revisit and does not retrieve wind direction from a single acquisition. CYGNSS uniquely combines global tropical coverage with $\sim$2.8\,h revisit, alleviating both spatial and temporal bottlenecks at the cost of being directionless and scalar. The fundamental challenge of reconstructing full 2D vector wind fields from purely scalar, directionless spaceborne observations---where the observation operator is formally non-invertible---thus remains unsolved for any globally deployable sensor, and it is precisely this challenge that the present work addresses.

Score-based diffusion models \cite{ho2020ddpm, song2021sde, karras2022edm} have gained widespread attention in recent years as a flexible generative framework, achieving excellent benchmark results in tasks like image synthesis, and are increasingly being applied to scientific problems. Particularly relevant to this paper is the score-based data assimilation paradigm \cite{rozet2023sda}, which provides a principled Bayesian method that combines a learned prior distribution $p(\mathbf{x})$ (encoded by a diffusion model) with observation likelihoods $p(\mathbf{y}|\mathbf{x})$ during the reverse sampling process. This framework has been extended to weather-scale data assimilation \cite{huang2024diffda, bao2024scoreda}, probabilistic forecasting \cite{price2024gencast}, and TC 3D structure reconstruction from sparse multi-variable observations \cite{han2026physics}. However, existing diffusion-based assimilation methods primarily deal with \emph{linear} observation operators (e.g., direct state observations on a coarse grid), where the mapping from states to observations is straightforward. The key case of \emph{nonlinear} observation operators---such as $H(u,v)=\sqrt{u^2+v^2}$, which maps 2D vector states to 1D scalar observations---introduces fundamental identifiability challenges that have not yet been explored in diffusion-based assimilation frameworks.

Three core challenges therefore remain unsolved: (1)~\emph{methodologically}, existing diffusion-based assimilation frameworks are formulated for linear observation operators and do not generalise to the nonlinear, many-to-one mapping $H(u,v)=\sqrt{u^2+v^2}$ characteristic of GNSS-R wind-speed retrievals; (2)~\emph{physically}, no learned prior can by itself lift the directional degeneracy of scalar observations, and additional boundary-layer structure must be injected to identify wind direction; (3)~\emph{observationally}, real CYGNSS data are not only directionless but also spatially sparse and saturating above $\sim$30\,m\,s$^{-1}$, demanding adaptive down-weighting of unreliable observations and robustness to irregular inner-core coverage. To address these challenges, we develop a physics-guided diffusion framework, hereafter QiFeng, that reconstructs---for the first time---global kilometre-scale TC inner-core vector wind fields from scalar satellite observations alone. A score-based diffusion model pre-trained offline on high-resolution Hurricane Weather Research and Forecasting (HWRF) simulations \cite{tallapragada2016hwrf} encodes a learned prior over physically reasonable $(u_{10}, v_{10})$ fields, and at inference each denoising step is guided by (i)~a \emph{nonlinear likelihood} that enforces consistency with CYGNSS via $H(u,v)=\sqrt{u^2+v^2}$; (ii)~three \emph{TC boundary-layer constraints}---low-level divergence \cite{holton2013dynamics}, bounded inflow angle ($10^\circ$--$35^\circ$) \cite{zhang2012inflow}, and positive cyclonic vorticity---injected as gradient guidance to resolve directional ambiguity. The method takes only CYGNSS scalar wind-speed observations as input; no concurrent NWP forecasts, HWRF outputs, or auxiliary wind products are required. The output is a spatially continuous $(u_{10}, v_{10})$ field on a $256\times 256$ grid covering $384\,\text{km}\times 384\,\text{km}$ at 1.5\,km resolution, in contrast to the existing CYGNSS L3 SCG product \cite{mayers2023scg} that delivers only along-track scalar wind speeds at $\sim$11\,km.

We apply QiFeng to 4{,}955 $\pm$3-h snapshots of 249 independent TCs (January 2020--September 2022) spanning all six active basins, including the Western North Pacific, Northern Indian Ocean and Southern Hemisphere where no routine aircraft reconnaissance exists, and validate against IBTrACS best-track \cite{knapp2010ibtracs}, ERA5, CCMP, SAR, dropsondes and Tail Doppler Radar (47 cases, $>$750{,}000 paired points). In controlled OSSEs with known ground truth, $V_{\max}$ MAE is 2.2--3.2\,m\,s$^{-1}$ and inflow-angle RMSE is $7^\circ$--$9^\circ$ across full-coverage, no-eye and eyewall-only observation geometries, outperforming parametric vortex and interpolation baselines. On the full sample, QiFeng reduces the systematic $V_{\max}$ bias against IBTrACS by $\sim$79\% and $\sim$75\% relative to ERA5 and CCMP, and a CYGNSS-intrinsic Observation Coverage Sufficiency (OCS) criterion, derived purely from satellite-observation metrics without external references, flags reliable snapshots and yields $V_{\max}$ correlation $R=0.80$ with IBTrACS on this subset. Ablation shows the boundary-layer constraints reduce wind-direction RMSE by $\sim$70\% on the OCS subset without degrading speed accuracy, and joint assimilation of only 11 dropsondes with sparse CYGNSS for TC FIONA (2022) cuts the cross-eye profile RMSE by 42\%, outlining a route to fuse CYGNSS with SFMR, SAR and scatterometer data in future operational TC analyses.

The remainder of this paper follows a Results--Discussion--Methods structure. Section~\ref{sec:results} presents OSSE and real-data experimental results. Section~\ref{sec:discussion} discusses major findings, methodological contributions, limitations, and future directions. Section~\ref{sec:method} details the framework, including data, nonlinear likelihood guidance, physical constraints, observation error models, evaluation metrics, and OSSE design.

\section{Results}
\label{sec:results}

Across 4,955 $\pm$3-h snapshots of 249 TCs from 2020 to 2022 (Fig.~\ref{fig:framework}d), the reconstructions capture the eye, eyewall, and asymmetric inner-core structure at 1.5\,km resolution in all six active basins, including the Western North Pacific, Northern Indian Ocean and Southern Hemisphere where no routine aircraft reconnaissance exists. The overall pipeline fuses a learned TC vector wind-field prior with sparse CYGNSS scalar observations and TC boundary-layer physical constraints during reverse diffusion sampling (Fig.~\ref{fig:framework}; full methodological details in Section~\ref{sec:method}). It proceeds in three stages: (i)~prior learning on HWRF simulations via EDM \cite{karras2022edm} (Fig.~\ref{fig:framework}a); (ii)~TC-centric CYGNSS observation collection with Observation Coverage Sufficiency (OCS) screening (Fig.~\ref{fig:framework}b); and (iii)~physics-guided nonlinear reverse diffusion sampling, in which a nonlinear likelihood term and three physical constraints (low-level divergence, bounded inflow angle, and positive cyclonic vorticity) jointly guide the reconstruction of the $(u_{10}, v_{10})$ field on a $256\times 256$ grid covering $384\,\text{km} \times 384\,\text{km}$ at 1.5\,km resolution (Fig.~\ref{fig:framework}c).

Section~\ref{sec:orbit-scenarios} benchmarks our approach against baseline methods in controlled OSSE environments with known ground truth; Section~\ref{sec:real-data} applies it to real CYGNSS observational data. Detailed definitions of data processing, model framework, and evaluation metrics are found in Sections~\ref{sec:data}, \ref{sec:nonlinear-likelihood-guidance}, and \ref{sec:metrics}, respectively.

\begin{figure}[p]
	\centering
	\includegraphics[width=\textwidth]{figs/arti.jpg}
\end{figure}

\begin{figure}[!t]
	\caption{Overview of the QiFeng framework and cross-basin TC coverage.
		\textbf{(a)} Score-based prior learning: a denoising network $D_\theta$ is trained on high-resolution HWRF TC simulations via the EDM framework, progressively denoising from Gaussian noise ($t=1$) to realistic TC wind field structures ($t=0$).
		\textbf{(b)} TC-centric domain extraction and OCS screening: CYGNSS specular points within a $\pm$3\,h window are collected into the TC-centric grid, and each snapshot is labeled by three computable criteria---observation count, spatial density, and azimuthal coverage (see Section~\ref{sec:coverage-criteria} for thresholds).
		\textbf{(c)} Physics-guided nonlinear reverse diffusion sampling: a nonlinear likelihood term matches denoising estimates with CYGNSS scalar observations via $H(u,v)=\sqrt{u^2+v^2}$, while three TC boundary layer constraints---low-level divergence, bounded inflow angle, and positive cyclonic vorticity---are injected as gradient guidance to resolve the directional ambiguity. The output is a 1.5\,km resolution $(u_{10}, v_{10})$ vector wind field.
		\textbf{(d)} Cross-basin TC coverage for this study (January 2020 to September 2022): 4,955 $\pm$3\,h snapshots of 249 TCs across six basins (NA, EP, WP, NI, SI, SP). Solid/semi-transparent dots indicate snapshots meeting/not meeting OCS criteria; bar chart shows sample size per basin (dark/light colors for OCS pass/fail).}
	\label{fig:framework}
\end{figure}

\subsection{OSSE Orbital Coverage Scenarios and Baseline Method Comparison}
\label{sec:orbit-scenarios}
To systematically evaluate QiFeng's reconstruction capability and robustness to different observation geometries in a controlled OSSE environment with known ground truth, we designed three typical CYGNSS orbital coverage scenarios (detailed in Section~\ref{sec:osse-design} and Supplementary Figure~S6), representing full inner-core coverage, no-eye coverage, and eyewall-only coverage, respectively. In this section, we quantitatively compare QiFeng against two classic baseline methods---cubic interpolation and the \cite{holland1980vortex} parametric vortex model---under the three orbital coverage scenarios in the OSSE controlled environment (experimental design in Section~\ref{sec:osse-design}). Evaluation metric definitions are found in Section~\ref{sec:metrics}. Cubic interpolation can only reconstruct scalar wind speed fields and cannot reconstruct wind direction information ($u$/$v$ components); therefore, inflow angle errors are not applicable (marked as ``---''). The Holland parametric model, while able to output complete vector wind fields, performs poorly in asymmetric TC structures due to its axisymmetric assumptions.

\textbf{Structural parameter accuracy} (Table~\ref{tab:baselines}).
QiFeng outperforms the two baseline methods across most structural parameters. Taking the most critical $V_{\max}$ as an example, under Scenario A, QiFeng's MAE is only $2.2$\,m\,s$^{-1}$, representing a reduction of approximately 69\% and 74\% compared to cubic interpolation and the Holland model, respectively. Consistent advantages are also achieved for RMW and wind radii (R34/R50/R64) (see Table~\ref{tab:baselines} for details). In the wind direction dimension, QiFeng's inflow angle RMSE is about $7^\circ$--$9^\circ$ across all scenarios, while the Holland model yields $11^\circ$--$13^\circ$ due to its fixed inflow angle parameterization. As coverage becomes sparser (from Scenario A to C), QiFeng's performance degradation is smaller than that of the baseline methods.

\begin{table}[htbp]
	\centering
	\caption{Comparison of structural parameters between QiFeng reconstruction and baseline methods (OSSE, physical constraint weight selection detailed in Section~\ref{sec:lambda-selection}, $n=366$).
		MAE, Bias (Reconstruction $-$ HWRF Ground Truth), and RMSE are reported separately.
		$N_\mathrm{ref}$ is the number of valid pairs (in parentheses).
		Cubic interpolation only outputs scalar wind speeds and cannot reconstruct wind direction, hence the inflow angle is marked as ``---''.
		\textbf{Bold} = best; \underline{underline} = second-best.}
	\label{tab:baselines}
	\setlength{\tabcolsep}{2pt}
	\resizebox{\textwidth}{!}{%
		\begin{tabular}{l l rrr rrr rrr rrr rrr r}
			\toprule
			& & \multicolumn{3}{c}{$V_{\max}$ (m\,s$^{-1}$)} & \multicolumn{3}{c}{RMW (km)} & \multicolumn{3}{c}{R34 (km)} & \multicolumn{3}{c}{R50 (km)} & \multicolumn{3}{c}{R64 (km)} & Inflow \\
			\cmidrule(lr){3-5}\cmidrule(lr){6-8}\cmidrule(lr){9-11}\cmidrule(lr){12-14}\cmidrule(lr){15-17}\cmidrule(lr){18-18}
			Scenario & Method
			& MAE & Bias & RMSE
			& MAE & Bias & RMSE
			& MAE ($N_\mathrm{ref}$) & Bias & RMSE
			& MAE ($N_\mathrm{ref}$) & Bias & RMSE
			& MAE ($N_\mathrm{ref}$) & Bias & RMSE
			& RMSE ($^\circ$) \\
			\midrule
			\multirow{3}{*}{\rotatebox{90}{\scriptsize A: Full}}
			& QiFeng & \textbf{2.2} & \tb{$-1.2$} & \textbf{3.1} & \textbf{8.0} & \tb{$+0.1$} & \textbf{13.9} & \textbf{4.3} (268) & \tb{$+0.4$} & \textbf{9.2} & \textbf{3.3} (154) & \tb{$-1.1$} & \textbf{5.5} & \textbf{2.2} (107) & \tb{$\pm 0.0$} & \textbf{4.0} & \textbf{7.0} \\
			& Cubic & \underline{7.1} & \underline{$+6.8$} & \underline{8.7} & 23.4 & $+12.7$ & 36.2 & 10.8 (265) & $+9.7$ & 19.1 & 15.5 (153) & $+15.2$ & 21.3 & 7.6 (107) & $+7.2$ & 9.8 & --- \\
			& Holland & 8.4 & $-8.4$ & 9.2 & \underline{13.4} & \underline{$-2.6$} & \underline{21.4} & \underline{5.3} (257) & \tb{$+0.4$} & \underline{9.3} & \underline{6.5} (150) & \underline{$+3.5$} & \underline{8.9} & \underline{5.3} (101) & \underline{$+0.9$} & \underline{7.4} & \underline{11.3} \\
			\midrule
			\multirow{3}{*}{\rotatebox{90}{\scriptsize B: No-Eye}}
			& QiFeng
			& \textbf{2.9} & \tb{$-1.5$} & \textbf{4.4}
			& \textbf{11.8} & \tb{$+2.8$} & \textbf{19.2}
			& \underline{4.5} (264) & \tb{$+1.0$} & \underline{9.8}
			& \textbf{3.4} (149) & \tb{$-0.5$} & \textbf{5.6}
			& \textbf{2.1} (102) & \tb{$\pm 0.0$} & \textbf{4.0}
			& \textbf{8.7} \\
			& Cubic
			& \underline{6.7} & $+6.0$ & \underline{8.5}
			& 27.6 & $+16.8$ & 39.7
			& 11.3 (263) & $+9.6$ & 20.1
			& 14.5 (156) & $+13.0$ & 20.7
			& 6.8 (110) & $+3.4$ & 9.6
			& --- \\
			& Holland
			& 10.3 & \underline{$-3.5$} & 16.7
			& \underline{15.9} & \underline{$-3.2$} & \underline{23.0}
			& \textbf{4.3} (259) & \underline{$+1.4$} & \textbf{7.8}
			& \underline{4.7} (153) & \underline{$+2.1$} & \underline{6.6}
			& \underline{4.3} (107) & \underline{$+0.7$} & \underline{5.8}
			& \underline{12.9} \\
			\midrule
			\multirow{3}{*}{\rotatebox{90}{\scriptsize C: Eyewall}}
			& QiFeng
			& \textbf{3.2} & \tb{$-1.7$} & \textbf{4.7}
			& \textbf{14.3} & \underline{$+1.6$} & \textbf{23.2}
			& \underline{8.3} (258) & \underline{$+3.3$} & \underline{16.0}
			& \textbf{6.4} (149) & \tb{$-0.4$} & \underline{9.7}
			& \textbf{4.2} (102) & \tb{$+1.1$} & \textbf{5.8}
			& \textbf{9.0} \\
			& Cubic
			& 11.2 & $+10.6$ & 14.9
			& 33.9 & $+11.5$ & 51.4
			& 23.6 (259) & $+20.8$ & 38.0
			& 42.2 (156) & $+40.3$ & 49.7
			& 35.3 (106) & $+31.8$ & 43.5
			& --- \\
			& Holland
			& \underline{9.3} & \underline{$-4.0$} & \underline{14.3}
			& \underline{16.8} & \tb{$-1.4$} & \underline{24.4}
			& \textbf{6.2} (260) & \tb{$+2.4$} & \textbf{12.2}
			& \underline{6.7} (153) & \underline{$+4.2$} & \textbf{9.1}
			& \underline{5.4} (108) & \underline{$+2.5$} & \underline{7.5}
			& \underline{13.0} \\
			\bottomrule
	\end{tabular}}
\end{table}

\begin{figure}[htbp]
	\centering
	\includegraphics[width=\textwidth]{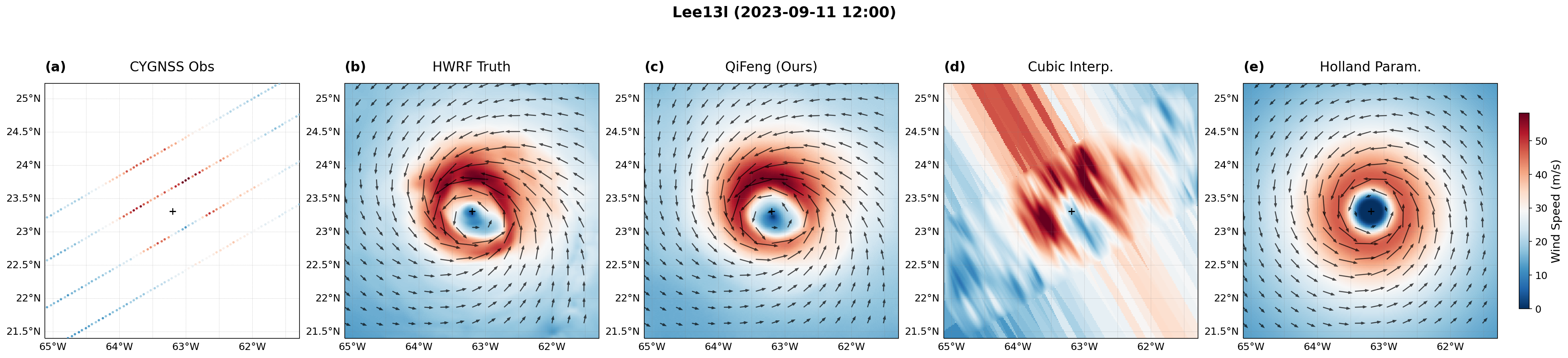}
	\caption{Spatial comparison of baseline methods for TC LEE (2023-09-11 12UTC) under OSSE Scenario B (no-eye coverage).
		(a) Simulated CYGNSS observations: three diagonal tracks crossing the TC domain, bypassing the eye (marked with ``+'');
		(b) HWRF ground truth field: mature TC structure, eyewall wind speeds exceeding 50\,m\,s$^{-1}$, wind speed distribution exhibiting significant asymmetry (northeast quadrant stronger than southwest quadrant);
		(c) QiFeng reconstruction: effectively captures the eye, asymmetric eyewall structure, and cyclonic wind directions, highly consistent with the ground truth;
		(d) Cubic interpolation: produces banding results only along tracks, failing to reconstruct the vortex structure and wind directions;
		(e) Holland parametric model: reconstructs an axisymmetric vortex but fails to reflect the asymmetric features in the ground truth.}
	\label{fig:osse-baselines-lee}
\end{figure}

Fig.~\ref{fig:osse-baselines-lee} uses TC LEE (2023-09-11 12UTC) to demonstrate the spatial reconstruction comparison under Scenario B (no-eye coverage). Despite the observation tracks bypassing the TC eye, QiFeng reconstructs the eye position and scale, the asymmetric eyewall structure (northeast quadrant stronger than southwest), and reasonable cyclonic wind directions, closely matching the HWRF ground truth (Fig.~\ref{fig:osse-baselines-lee}b,c). In contrast, cubic interpolation produces banding artifacts along observation tracks and entirely fails to reconstruct the vortex structure or wind directions (Fig.~\ref{fig:osse-baselines-lee}d), while the Holland parametric model reconstructs a basic eye-eyewall structure but imposes an unrealistic axisymmetric wind speed distribution, failing to capture the significant asymmetric features present in the ground truth (Fig.~\ref{fig:osse-baselines-lee}e). This case demonstrates that under sparse conditions where observations do not cover the eye, QiFeng can still rely on the diffusion model prior to reconstruct the TC's asymmetric fine structures---a capability that traditional interpolation methods (due to the lack of physical priors) and parametric models (due to axisymmetric assumptions) cannot achieve.

\subsection{TC Vector Wind Field Reconstruction and Validation Based on Real CYGNSS Observations}
\label{sec:real-data}

After evaluating QiFeng's performance in the controlled OSSE environment, this section applies it to real CYGNSS observational data from January 2020 to September 2022 (data processing workflow detailed in Section~\ref{sec:data}) for systematic multidimensional evaluation. Physical constraint weights and other hyperparameter selections are detailed in Sections~\ref{sec:lambda-selection} and \ref{sec:guided-sampling}, and evaluation metric definitions in Section~\ref{sec:metrics}.

This section first uses spatial comparisons of three typical TC cases with high-resolution SAR wind fields to intuitively demonstrate QiFeng's capability in characterizing TC inner-core structures (Section~\ref{sec:val-sar}). Due to the varying spatiotemporal coverage quality of real CYGNSS observations, Section~\ref{sec:coverage-criteria} establishes an OCS criterion. Subsequently, in the full-scale statistical evaluation over 4,955 $\pm$3\,h TC snapshots (Section~\ref{sec:struct-table}), performance metrics are reported for both the full sample and the subset meeting the OCS criterion. Sections~\ref{sec:val-dropsonde} and \ref{sec:val-tdr} utilize dropsonde and TDR airborne observations, respectively, for independent spatiotemporal matching validation on the full reconstructed results (matching methods in Section~\ref{sec:metrics}), and demonstrate the performance differences before and after applying the OCS criterion. Section~\ref{sec:fusion-case-study} takes TC FIONA as an example to demonstrate the potential of multi-source observation fusion. Ensemble uncertainty estimation, a reconstruction failure case analysis, and the physical constraint weight $\lambda$ sensitivity analysis are provided in the Supplementary Information (Sections~S5--S7).

\subsubsection{Typical TC Cases and SAR Comparison}
\label{sec:val-sar}

To further intuitively demonstrate QiFeng's spatial structure reconstruction capabilities, we selected three typical cases: TC IAN (2022, an Atlantic rapid intensification TC), TC HINNAMNOR (2022, the strongest TC of the year in the Western North Pacific), and TC EMNATI (2022, a strong Southern Hemisphere TC in the Southwest Indian Ocean), comparing QiFeng-reconstructed wind fields with near-simultaneous spaceborne SAR (resolution $\sim$3\,km) observed wind speeds. EMNATI, being located in the Southern Hemisphere with a cyclonic circulation direction opposite to the Northern Hemisphere (clockwise rotation), can be used to verify QiFeng's adaptability to Southern Hemisphere TCs.

A time difference exists between the SAR imaging moment and the CYGNSS observation time corresponding to the QiFeng reconstruction. Specifically, the time offsets are approximately 31 minutes for TC IAN (SAR: 2022-09-27 23:28:46 UTC; QiFeng: 2022-09-28 00:00 UTC), 2 hours 49 minutes for TC HINNAMNOR (SAR: 2022-08-30 21:11:23 UTC; QiFeng: 2022-08-31 00:00 UTC), and 2 hours 4 minutes for TC EMNATI (SAR: 2022-02-22 02:03:34 UTC; QiFeng: 2022-02-22 00:00 UTC). Because the TC translates during this interval, the SAR-observed eye position deviates from the IBTrACS-reported center. Maximum-contrast eye localization on the SAR wind fields (the extremum point of the difference between 100\,km and 10\,km uniform filtering) yields SAR eye offsets of approximately $14.5$\,km for IAN, $74.5$\,km for HINNAMNOR (primarily an east-west shift of $72.8$\,km, reflecting its rapid movement), and $63.2$\,km for EMNATI (primarily an east-west shift of $57.2$\,km). In the figures, spatial wind fields (Panel a) retain the original pixel positions of each data source without offsets: QiFeng and CCMP are centered on the IBTrACS reported center, while the SAR wind field is centered on its own localized eye. For the West-East (W-E) cross-eye profiles (Panel b), each data source extracts the cross-eye section along its respective eye latitude, using its respective eye longitude as the zero point, enabling structural comparisons in relative radial coordinates. Because TC structural evolution within the time difference cannot be eliminated, the comparison results should be viewed as approximate rather than strictly synchronous.

\begin{figure}[htbp]
	\centering
	\includegraphics[width=\textwidth]{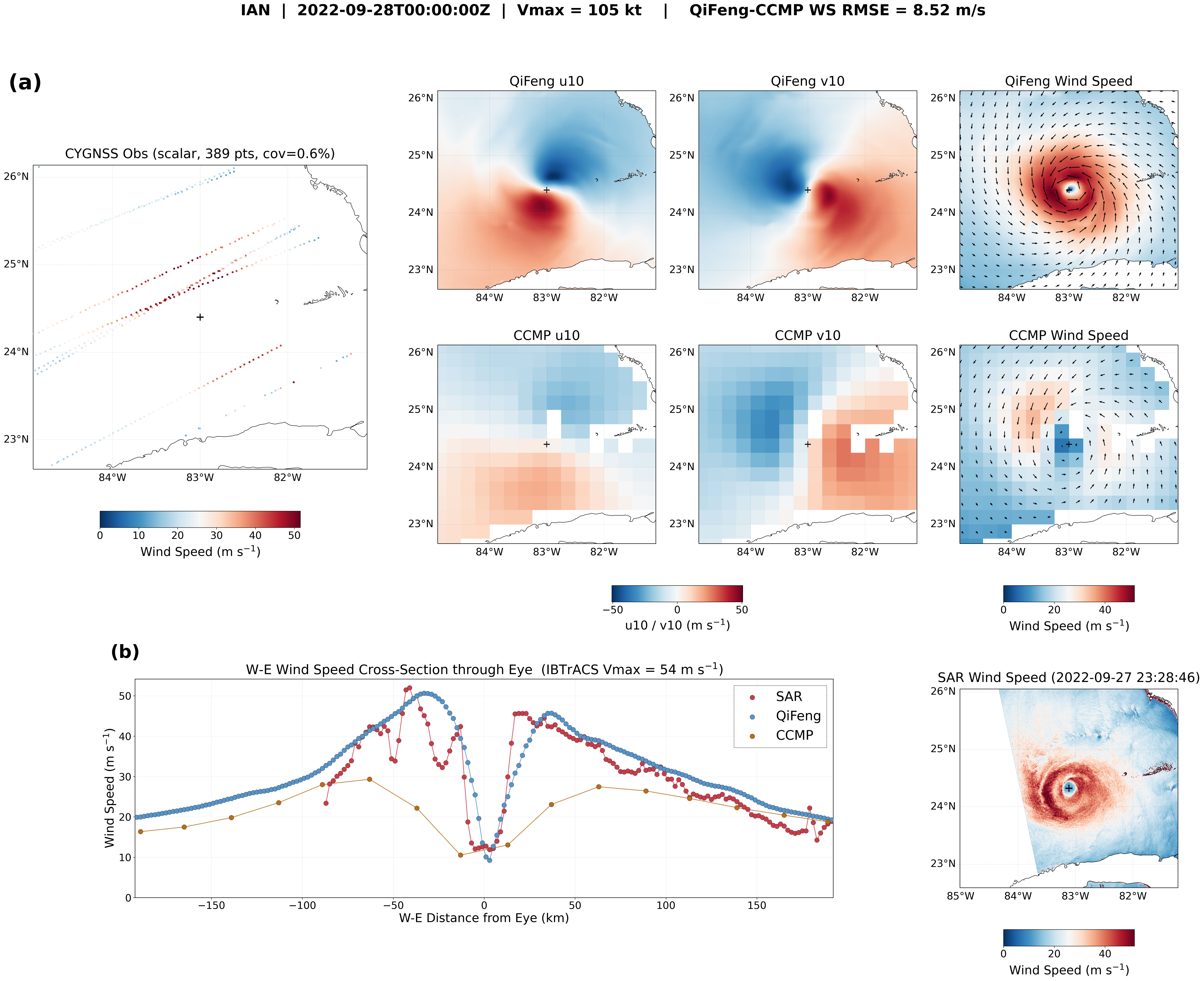}
	\caption{Comparison between QiFeng reconstruction and SAR wind field for TC IAN (2022-09-28 00UTC, $V_{\max} = 105$\,kt). The SAR eye is about $14.5$\,km from the IBTrACS center.
		(a) Top: CYGNSS observations (389 pts, cov=0.6\%), QiFeng reconstructed $u_{10}$/$v_{10}$ components and vector wind speed field, CCMP $u_{10}$/$v_{10}$ and wind speed field;
		(b) Bottom left: West-East cross-eye wind speed profile comparison (SAR red, QiFeng blue, CCMP yellow); Bottom right: SAR observed wind speed field (imaging time 2022-09-27 23:28:46 UTC, about 31 minutes difference from QiFeng time).}
	\label{fig:sar-ian}
\end{figure}

Fig.~\ref{fig:sar-ian} shows the comparison for TC IAN (2022-09-28 00UTC, IBTrACS $V_{\max} = 105$\,kt). In this case, the time difference is only about 31 minutes, the SAR eye offset is only $14.5$\,km, and the spatial structure comparability is good. QiFeng reconstructed an identifiable TC eye and eyewall structure, with the eyewall's maximum wind speed distribution exhibiting asymmetric features, qualitatively consistent with the wind field structure observed by SAR. In the W-E cross-eye profile, QiFeng (blue) roughly tracks the bimodal structure of SAR (red) in the inner core region ($|r| < 100$\,km): the west eyewall peak is approximately 50.6 for QiFeng vs. 52.0\,m\,s$^{-1}$ for SAR, and the east eyewall peak is approximately 45.7 for QiFeng vs. 45.6\,m\,s$^{-1}$ for SAR; the lowest wind speed near the eye ($|r| < 20$\,km) is about 9.3 for QiFeng vs. 11.9\,m\,s$^{-1}$ for SAR. A bin-by-bin quantitative comparison of the W-E profile within the SAR's effective coverage range ($[-87, 191]$\,km, a total of 140 2\,km radial bins) yields an overall RMSE of 5.52\,m\,s$^{-1}$, MAE of 4.14\,m\,s$^{-1}$, Bias of $+2.89$\,m\,s$^{-1}$, and correlation $R = 0.892$; the inner-core region RMSE is 6.28\,m\,s$^{-1}$ ($R = 0.815$), while the outer region RMSE drops to 3.47\,m\,s$^{-1}$ ($R = 0.947$). In contrast, limited by its $0.25^\circ$ resolution, CCMP severely underestimates wind speeds across the entire profile (profile RMSE of 18.57\,m\,s$^{-1}$, Bias of $-13.58$\,m\,s$^{-1}$), with a peak wind speed of only about 29\,m\,s$^{-1}$, less than 55\% of the IBTrACS reported value (105\,kt $\approx$ 54\,m\,s$^{-1}$), failing to capture the inner-core structural features.

\begin{figure}[htbp]
	\centering
	\includegraphics[width=\textwidth]{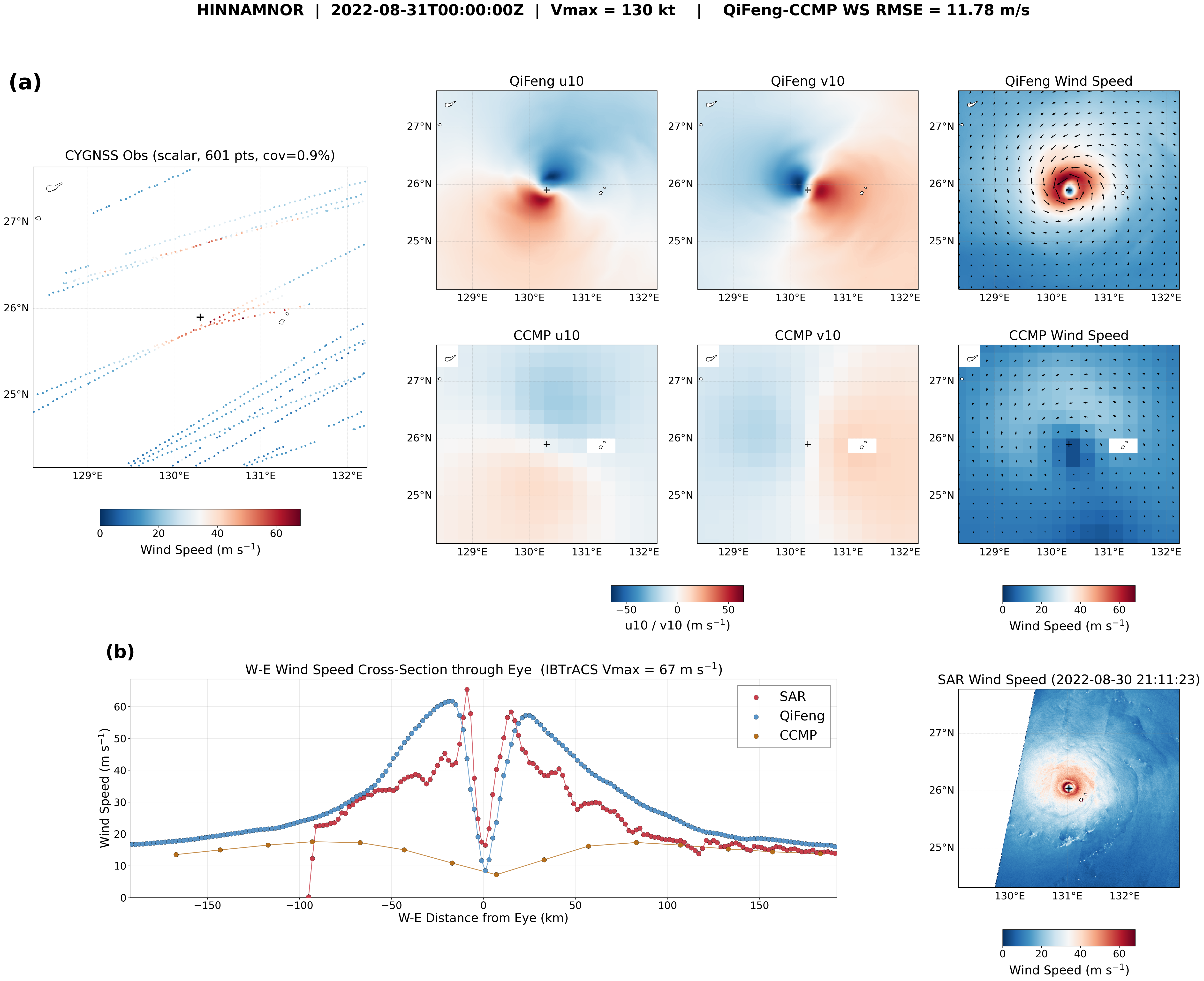}
	\caption{Comparison between QiFeng reconstruction and SAR wind field for TC HINNAMNOR (2022-08-31 00UTC, $V_{\max} = 130$\,kt). The SAR eye is about $74.5$\,km from the IBTrACS center (primarily an east-west offset of $72.8$\,km).
		(a) Top: CYGNSS observations (601 pts, cov=0.9\%), QiFeng reconstructed $u_{10}$/$v_{10}$ components and vector wind speed field, CCMP $u_{10}$/$v_{10}$ and wind speed field;
		(b) Bottom left: West-East cross-eye wind speed profile comparison (SAR red, QiFeng blue, CCMP yellow); Bottom right: SAR observed wind speed field (imaging time 2022-08-30 21:11:23 UTC, about 2 hours 49 minutes difference from QiFeng time).}
	\label{fig:sar-hinnamnor}
\end{figure}

Fig.~\ref{fig:sar-hinnamnor} shows the comparison for TC HINNAMNOR (2022-08-31 00UTC, IBTrACS $V_{\max} = 130$\,kt). HINNAMNOR was one of the strongest TCs in the Western North Pacific in 2022, and at this time was undergoing rapid intensification, with CYGNSS providing 601 observation points (0.9\% coverage). Despite a time difference of about 2 hours and 49 minutes between SAR and QiFeng, QiFeng's reconstructed wind field still exhibits a compact eye structure and steep eyewall wind speed gradients. The W-E cross-eye profile shows that QiFeng's west eyewall peak is about 61.7\,m\,s$^{-1}$, close to the SAR peak ($\sim$65.3\,m\,s$^{-1}$, a difference of $-3.6$\,m\,s$^{-1}$), while the east eyewall peak is about 57.2 for QiFeng vs. 58.3\,m\,s$^{-1}$ for SAR, with the bimodal radial positions being basically consistent. Near the eye ($|r| < 20$\,km), both observed a sudden drop in wind speed, with SAR down to $\sim$16.5\,m\,s$^{-1}$ and QiFeng down to $\sim$8.5\,m\,s$^{-1}$, a difference of $-8.0$\,m\,s$^{-1}$, reflecting smoothing effects in extreme wind speed gradient regions. Within the SAR coverage range (144 radial bins), the overall RMSE is 9.75\,m\,s$^{-1}$, MAE is 7.97\,m\,s$^{-1}$, Bias is $+5.60$\,m\,s$^{-1}$, and $R = 0.830$; the inner-core region ($|r| < 100$\,km) RMSE is 11.52\,m\,s$^{-1}$ ($R = 0.696$), and the outer region RMSE drops to 3.85\,m\,s$^{-1}$ ($R = 0.729$). HINNAMNOR's larger time difference ($\sim$2 hours 49 minutes) leads to overall quantitative metrics slightly inferior to the IAN case. CCMP underestimates even more significantly in this strong TC case (profile RMSE 20.11\,m\,s$^{-1}$, Bias $-14.33$\,m\,s$^{-1}$), with a maximum wind speed of only $\sim$18\,m\,s$^{-1}$, less than 27\% of the IBTrACS reported value (130\,kt $\approx$ 66.9\,m\,s$^{-1}$).

A Southern Hemisphere case, TC EMNATI (2022-02-22 00UTC, $V_{\max} = 90$\,kt, Southwest Indian Ocean, clockwise cyclonic circulation), further verifies QiFeng's cross-hemisphere adaptability. Despite a $\sim$2\,h time difference with SAR and partial SAR coverage, QiFeng reconstructs a clear TC eye and asymmetric eyewall (W-E profile RMSE = 6.18\,m\,s$^{-1}$, $R = 0.758$), confirming the effectiveness of the coordinate-flipping strategy for Southern Hemisphere TCs (Supplementary Figure~S1). CCMP's peak wind speed was about 33.9\,m\,s$^{-1}$, only 73\% of the IBTrACS reported value (90\,kt $\approx$ 46.3\,m\,s$^{-1}$). This case demonstrates that QiFeng can effectively handle the circulation reversal characteristic of Southern Hemisphere TCs and provide wind field reconstructions with inner-core resolving capability in basins lacking routine aircraft reconnaissance.

\subsubsection{Observation Coverage Sufficiency (OCS) Criterion}
\label{sec:coverage-criteria}

The SAR cases above demonstrated QiFeng's strong performance in TC wind field reconstruction, but real CYGNSS observations, constrained by satellite orbits and sea states, frequently exhibit sparse observation points or fail to fully capture TC inner-core structures. When CYGNSS observations lack spatial diversity (e.g., only a single track that fails to sample the TC core region), QiFeng cannot reconstruct a complete vortex structure from limited scalar information. The TC TEDDY case (Supplementary Figure~S5) is a typical reconstruction failure mode, where only 65 observation points distributed along a single track caused QiFeng to erroneously reconstruct the high wind speed signals into a frontal-like banded flow pattern rather than a closed vortex. This indicates that QiFeng requires a certain degree of spatial observation coverage over key TC structures (especially the eyewall region) to effectively guide the reconstruction. Indiscriminately applying QiFeng to all observation samples could introduce large uncertainties, so an OCS criterion is needed to determine whether current observation conditions are suitable for high-quality QiFeng reconstruction.

\begin{figure}[htbp]
	\centering
	\includegraphics[width=\textwidth]{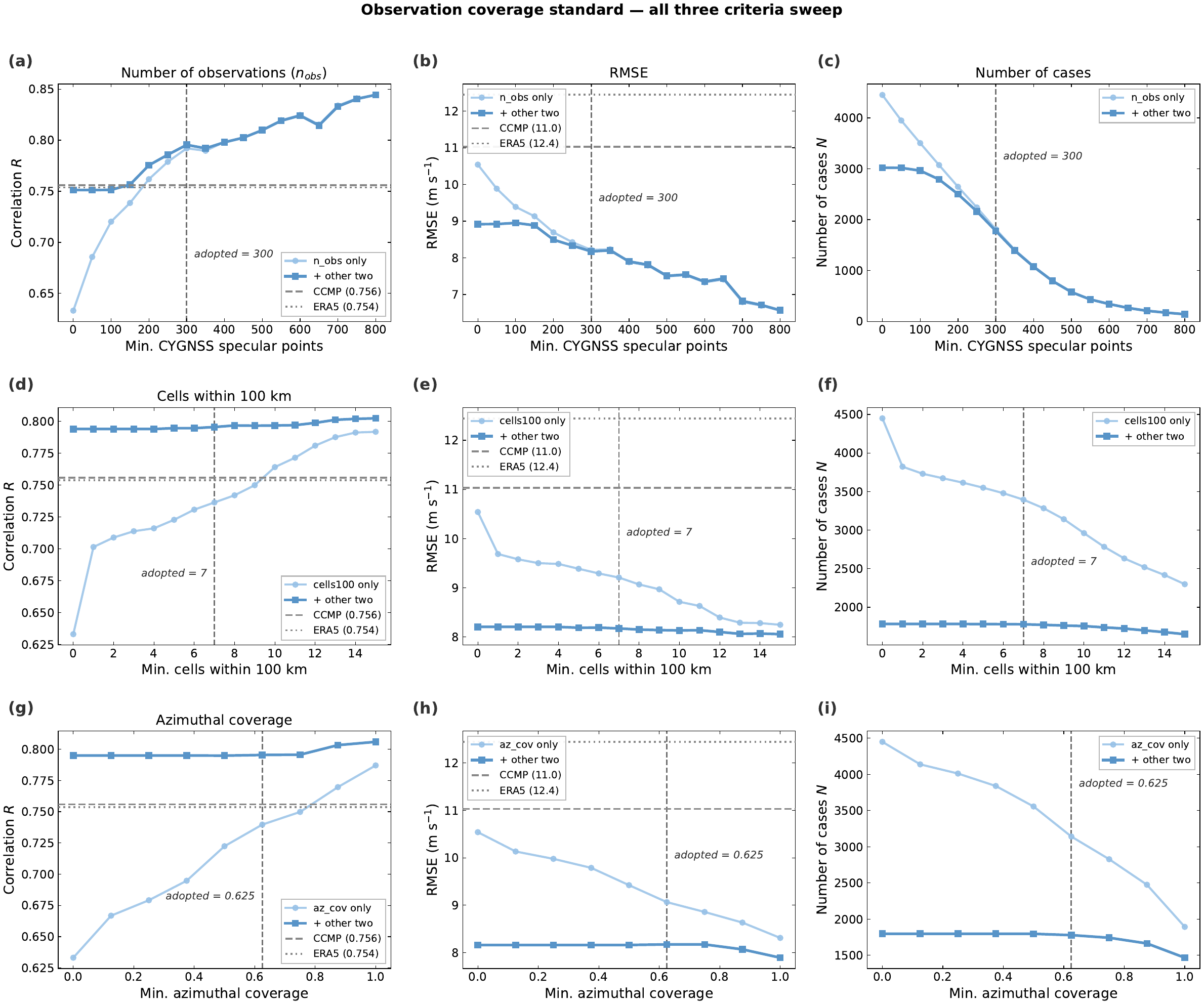}
	\caption{Threshold sensitivity analysis of the three OCS criteria (January 2020 to September 2022, 4,450 valid cases).
		Each row corresponds to a criterion: (a--c) number of observations $n_\mathrm{obs}$; (d--f) number of super-grid cells within 100\,km (cells100); (g--i) azimuthal coverage rate az\_cov.
		Each column displays correlation $R$ (left), RMSE (middle), and retained case count $N$ (right).
		Light-colored curves represent results applying only that single criterion, while dark-colored curves represent results when fixing the other two criteria at their adopted values.
		Gray dashed and dotted lines indicate CCMP and ERA5 full-sample baselines, respectively.
		Vertical dashed lines denote the adopted thresholds for each criterion ($n_\mathrm{obs} \geq 300$, cells100 $\geq 7$, az\_cov $\geq 0.625$).}
	\label{fig:threshold-sensitivity-all3}
\end{figure}

Fig.~\ref{fig:threshold-sensitivity-all3} presents a threshold sensitivity analysis for the three criteria: number of observations $n_\mathrm{obs}$, number of super-grid cells within 100\,km (cells100), and azimuthal coverage rate az\_cov. $n_\mathrm{obs}$ is the decisive screening criterion: $R$ shows a distinct inflection point in the $n_\mathrm{obs} = 200$--$300$ interval, leaping from $\sim$0.75 to $\sim$0.80, with RMSE simultaneously dropping from $\sim$9.0 to $\sim$8.1\,m\,s$^{-1}$. Cells100 and az\_cov also show significant screening effects when applied individually, but their marginal contribution becomes near zero (conditional pass rate $>$98\%) once $n_\mathrm{obs} \geq 300$ is met. However, they are retained as safety nets for spatial and azimuthal coverage. The joint application of the three criteria retained 1,779 cases (40.0\%) with $R = 0.795$ and RMSE $= 8.2$\,m\,s$^{-1}$, surpassing the full-sample baselines of CCMP ($R = 0.756$) and ERA5 ($R = 0.754$) (note that this compares a filtered subset with full-sample baselines).

Additional analysis confirms that QiFeng's structural parameter errors decrease with increasing effective observation coverage (Supplementary Figure~S2).

\begin{figure}[htbp]
	\centering
	\includegraphics[width=\textwidth]{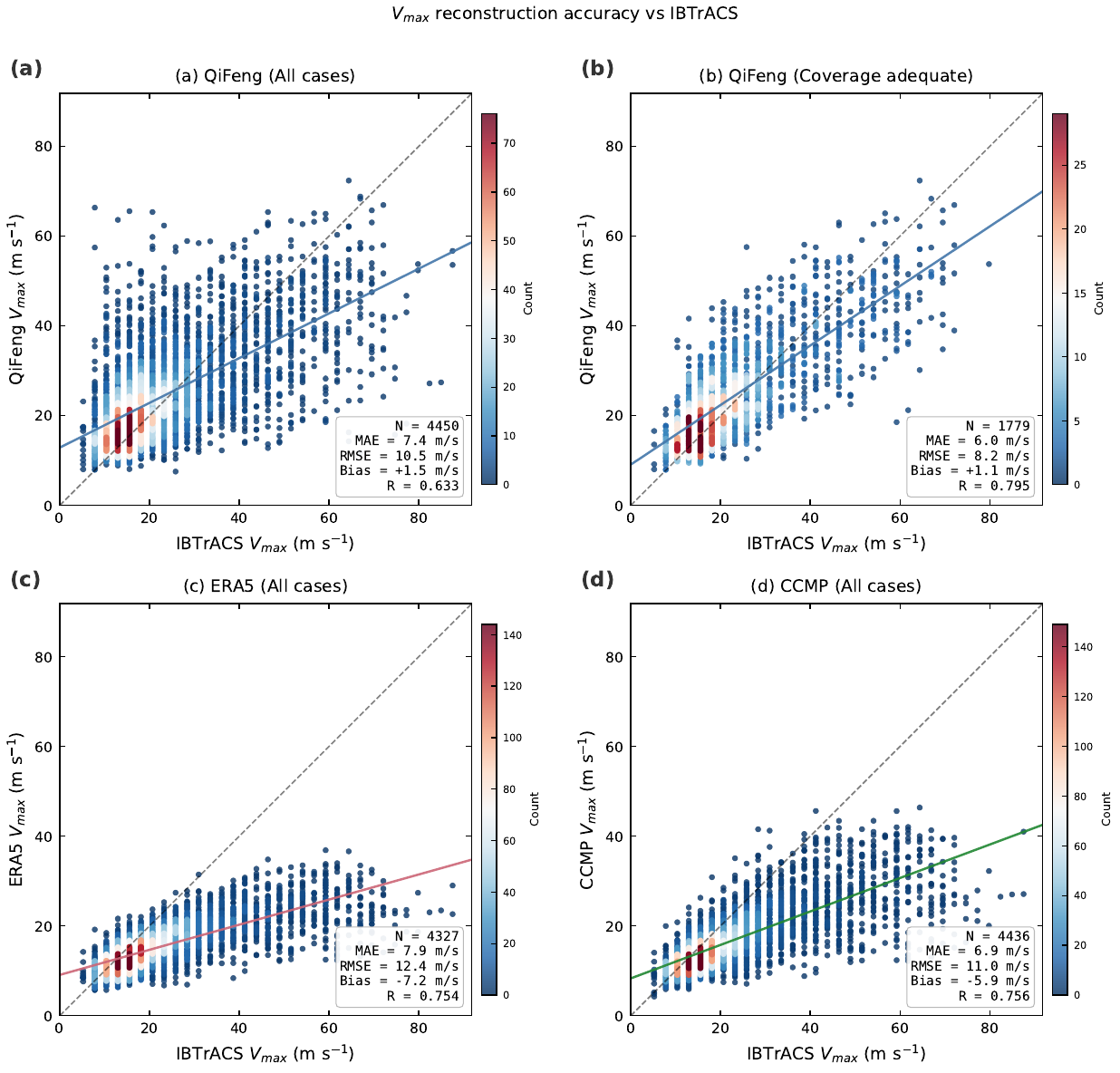}
	\caption{$V_{\max}$ scatter plot comparison against IBTrACS best-track $V_{\max}$.
		(a) QiFeng, all cases; (b) QiFeng, OCS-qualified (coverage adequate) subset; (c) ERA5 reanalysis, all cases; (d) CCMP satellite wind field, all cases.
		Colors indicate point density; gray dashed lines are 1:1 reference lines; solid lines are linear regression fits.
		QiFeng exhibits better consistency in high wind speed regimes after OCS screening, while ERA5 and CCMP show distinct systematic underestimations.}
	\label{fig:vmax-triple-scatter}
\end{figure}

Fig.~\ref{fig:vmax-triple-scatter} displays a scatter plot comparison among QiFeng, ERA5, and CCMP against IBTrACS best-track data for $V_{\max}$. Applying the OCS criterion, QiFeng demonstrates better consistency with IBTrACS in high wind speed ranges than ERA5 and CCMP, reducing the systematic intensity underestimation inherent in traditional reanalyses and satellite-merged products. This OCS criterion is consistently supported in subsequent independent validations using dropsondes (Section~\ref{sec:val-dropsonde}, Table~\ref{tab:dropsonde-metrics}) and TDR (Section~\ref{sec:val-tdr}, Table~\ref{tab:tdr-metrics}).

\subsubsection{Full-Sample Statistical Evaluation}
\label{sec:struct-table}

Tables~\ref{tab:struct-errors} and \ref{tab:basin-intensity} present quantitative evaluations of QiFeng's $(u,v)$ wind field reconstructions against IBTrACS best-track values and ERA5/CCMP reference wind fields across structural parameters and basin/intensity stratification. Table~\ref{tab:struct-errors} reports metrics for both the full sample (4,955 $\pm$3\,h TC snapshots, Jan 2020 to Sep 2022) and the OCS-qualified subset (1,960 snapshots). Table~\ref{tab:basin-intensity} provides basin and intensity-stratified statistics on the OCS-qualified subset (1,779 valid $V_{\max}$ pairs). Wind direction fidelity against ERA5/CCMP is reported in Supplementary Table~S1.

Regarding structural metrics (Table~\ref{tab:struct-errors}), three wind field sources---CYGNSS-based QiFeng reconstruction, ERA5 reanalysis ($0.25^\circ$), and CCMP v3.1 satellite-derived wind field ($0.25^\circ$)---were compared with IBTrACS. Before calculating metrics, ERA5 and CCMP's native $0.25^\circ$ $(u,v)$ fields were bilinearly interpolated onto the QiFeng 256$\times$256 grid (1.5\,km) to extract structural parameters on a unified grid. Note that this interpolation is an up-sampling operation, generating values bounded by the convex hull of original grid points, introducing no additional smoothing or $V_{\max}$ reduction---the $V_{\max}$ underestimation in ERA5 and CCMP stems from their native $\sim$25\,km resolution's inherent limitation in resolving TC inner cores, not from interpolation artifacts. All structural parameters (RMW, R34, R50, R64) were calculated as averages of four quadrants (NE/NW/SW/SE), consistent with the definition of IBTrACS USA\_RMW/R34/R50/R64; RMW is measured from the TC eye center (local wind speed minimum). To eliminate confounding effects of varying eye locations among data sources, all wind fields were first aligned to the grid center before calculating structural parameters. QiFeng's mean eye offset was $\sim$14.5\,km (median 5.4\,km). QiFeng's $V_{\max}$ bias is only $+1.5$\,m\,s$^{-1}$, reducing systematic underestimation magnitude by approximately 79\% and 75\% relative to ERA5 ($-7.2$\,m\,s$^{-1}$) and CCMP ($-5.9$\,m\,s$^{-1}$), respectively. For RMW, both ERA5 and CCMP exhibit a $\sim$+50\,km positive bias, reflecting the inherent resolving limits of $0.25^\circ$ resolution products, while QiFeng's bias is only $-7.1$\,km. Note that although CCMP's $V_{\max}$ MAE is slightly lower than QiFeng's (6.9 vs. 7.4\,m\,s$^{-1}$), its systematic negative bias indicates that the lower MAE masks significant intensity underestimation. After applying the OCS criterion (1,779 valid $V_{\max}$ pairs out of 1,960 OCS-qualified snapshots), the $V_{\max}$ RMSE decreased from 10.5 to 8.2\,m\,s$^{-1}$ (a 22\% reduction), showing that improving CYGNSS observation coverage quality directly yields systematic improvements in reconstruction accuracy (full metrics in Table~\ref{tab:struct-errors}).

To evaluate QiFeng's applicability across different basins and TC intensities, Table~\ref{tab:basin-intensity} provides stratified statistics on the OCS-qualified subset. Panel (a) reports QiFeng's $V_{\max}$ errors relative to IBTrACS categorized by WMO basins. The dataset covers 235 independent TCs across all six major TC basins: four Northern Hemisphere basins (NA 382 snapshots/68 TCs, EP 372/48, WP 449/61, NI 45/11) and two Southern Hemisphere basins (SI 366/44, SP 165/23), with Southern Hemisphere snapshots accounting for $\sim$30\%. QiFeng achieved a correlation $R > 0.73$ across all basins, performing best in NA and WP ($R > 0.84$), and relatively lower in Southern Hemisphere basins ($R \approx 0.74$--$0.76$). Bias structures exhibit basin differences: NA, possessing more strong TCs, shows a negative bias, reflecting an inner-core underestimation trend; WP, SI, and SP show positive biases ($+2.5$--$3.1$\,m\,s$^{-1}$), possibly relating to the diffusion prior's climatological mean bias when weak TCs are highly represented (detailed metrics by basin in Table~\ref{tab:basin-intensity}a). Consistency across basins indicates that the QiFeng framework---including the strategy of training on HWRF Northern Hemisphere simulations and handling Southern Hemisphere TCs uniformly via coordinate flipping---is robust across all basins.

Panel (b) stratifies by Saffir-Simpson intensity category, simultaneously comparing QiFeng, ERA5, and CCMP on the same OCS subset. All three wind field sources are evaluated on this identical subset, which was screened based on spatial coverage attributes of CYGNSS observations (input-side metrics) rather than reconstruction quality (output-side metrics), making the comparison fair and applicable to ERA5 and CCMP. Results reveal a significant intensity dependence: for weak TCs (TD/TS), ERA5 and CCMP perform comparably to or even better than QiFeng because their $0.25^\circ$ resolution is sufficient to reasonably characterize inner-core structures; but as TC intensity increases, the divergence among the three drastically widens---for Cat\,4--5, ERA5 and CCMP $V_{\max}$ RMSEs reach 40.3 and 36.1\,m\,s$^{-1}$, respectively, whereas QiFeng's is 18.7\,m\,s$^{-1}$, a 54\% reduction compared to ERA5 (Table~\ref{tab:basin-intensity}b). QiFeng's positive bias for weak TCs primarily stems from the diffusion prior's tendency to generate climatological mean structures higher than actual TD intensities; the negative bias for strong TCs is limited by CYGNSS L-band signal sensitivity saturation at high wind speeds.

Regarding wind direction fidelity, QiFeng's median wind direction RMSE against ERA5 and CCMP is $\sim$14$^\circ$ (full sample) and $\sim$11$^\circ$ (OCS-qualified), both within the natural variability range of TC boundary layer inflows ($15^\circ$--$30^\circ$). A near-zero directional bias ($\sim$$-2^\circ$--$-3^\circ$) confirms no systematic rotational shift. Independent TDR-based wind direction validation is presented in Section~\ref{sec:val-tdr}. Full metrics are provided in Supplementary Table~S1.

\begin{table}[htbp]
	\caption{MAE, Bias (Estimate $-$ IBTrACS), and RMSE for TC structural parameters, comparing QiFeng, ERA5, and CCMP against IBTrACS (4,955 snapshots, Jan 2020--Sep 2022).
		Units: $V_{\max}$ in m\,s$^{-1}$; RMW, R34, R50, R64 in km.
		$N_\mathrm{ref}$ = valid IBTrACS pairs (in parentheses).
		$^\dagger$R34 excludes IBT R34 $>$ 192\,km.
		``--'' = $<$3 valid pairs.
		$^\ddagger$OCS-qualified subset (Section~\ref{sec:coverage-criteria}).
		\textbf{Bold} = best; \underline{underline} = second-best.}
	\label{tab:struct-errors}
	\centering
	\setlength{\tabcolsep}{2pt}
	\resizebox{\textwidth}{!}{%
		\begin{tabular}{l r rrr rrr rrr rrr rrr}
			\toprule
			& & \multicolumn{3}{c}{$V_{\max}$ (m\,s$^{-1}$)} & \multicolumn{3}{c}{RMW (km)} & \multicolumn{3}{c}{R34$^\dagger$ (km)} & \multicolumn{3}{c}{R50 (km)} & \multicolumn{3}{c}{R64 (km)} \\
			\cmidrule(lr){3-5}\cmidrule(lr){6-8}\cmidrule(lr){9-11}\cmidrule(lr){12-14}\cmidrule(lr){15-17}
			Source & $N$ & MAE ($N_\mathrm{ref}$) & Bias & RMSE
			& MAE ($N_\mathrm{ref}$) & Bias & RMSE
			& MAE ($N_\mathrm{ref}$) & Bias & RMSE
			& MAE ($N_\mathrm{ref}$) & Bias & RMSE
			& MAE ($N_\mathrm{ref}$) & Bias & RMSE \\
			\midrule
			QiFeng & 4955
			& 7.4 (4450) & \underline{$+1.5$} & \underline{10.5}
			& \underline{39.4} (4342) & \underline{$-7.1$} & \underline{52.7}
			& \underline{65.1} (1404) & \underline{$-52.3$} & \underline{86.8}
			& 29.1 (568) & \underline{$-9.9$} & 38.1
			& 22.3 (271) & \underline{$+8.6$} & 29.4 \\
			QiFeng$^\ddagger$ & 1960
			& \textbf{6.0} (1779) & \tb{$+1.1$} & \textbf{8.2}
			& \textbf{35.8} (1750) & \tb{$-4.6$} & \textbf{48.7}
			& \textbf{36.8} (367) & \tb{$-15.7$} & \textbf{46.5}
			& \underline{28.0} (313) & \tb{$-7.9$} & \underline{36.8}
			& \underline{22.2} (159) & $+10.5$ & \underline{29.3} \\
			\midrule
			ERA5 & 4955
			& 7.9 (4327) & $-7.2$ & 12.4
			& 60.3 (4221) & $+50.7$ & 69.6
			& 86.4 (559) & $-85.0$ & 104.3
			& 46.5 (49) & $-46.5$ & 50.8
			& \multicolumn{3}{c}{--} \\
			CCMP & 4955
			& \underline{6.9} (4436) & $-5.9$ & 11.0
			& 59.0 (4328) & $+49.3$ & 68.3
			& 73.1 (703) & $-70.8$ & 93.8
			& \textbf{24.1} (163) & $-12.4$ & \textbf{30.7}
			& \textbf{16.4} (27) & \tb{$+4.4$} & \textbf{20.8} \\
			\bottomrule
	\end{tabular}}
\end{table}

\begin{table}[htbp]
	\caption{Basin-stratified (a) and intensity-stratified (b) statistics of QiFeng reconstructed $V_{\max}$ on the OCS-qualified subset ($N=1,779$ valid $V_{\max}$ pairs).
		(a) Categorized by WMO basin (based on TC center latitude/longitude), reporting only QiFeng reconstruction metrics;
		(b) Categorized by Saffir-Simpson intensity scale (based on IBTrACS $V_{\max}$), comparing QiFeng, ERA5, and CCMP on the same subset (screened using CYGNSS observation coverage metrics, equally applicable to all three sources).
		ERA5 and CCMP's $V_{\max}$ are extracted as maximum wind speeds after interpolating their native $0.25^\circ$ grids onto the QiFeng 256$\times$256 grid.
		Units: MAE, Bias (Estimate $-$ IBTrACS), and RMSE in m\,s$^{-1}$.
		Note: some TCs traverse multiple basins and are counted in each basin they enter; hence the sum of per-basin $N_\mathrm{TC}$ exceeds the total of 235 unique TCs.}
	\label{tab:basin-intensity}
	\centering
	\setlength{\tabcolsep}{4pt}
	
	\textbf{(a) Basin-stratified}\\[4pt]
	\begin{tabular}{l r r rrrr}
		\toprule
		Basin & $N_\mathrm{snap}$ & $N_\mathrm{TC}$ & MAE & Bias & RMSE & $R$ \\
		\midrule
		NA & 382 & 68 & 5.5 & $-2.9$ & 7.6 & 0.855 \\
		EP & 372 & 48 & 4.9 & $+0.5$ & 6.9 & 0.771 \\
		WP & 449 & 61 & 6.7 & $+3.1$ & 8.8 & 0.845 \\
		NI & 45 & 11 & 5.9 & $+0.1$ & 8.6 & 0.823 \\
		SI & 366 & 44 & 6.8 & $+2.8$ & 8.9 & 0.736 \\
		SP & 165 & 23 & 6.3 & $+2.5$ & 8.3 & 0.758 \\
		\midrule
		All & 1779 & 235 & 6.0 & $+1.1$ & 8.2 & 0.795 \\
		\bottomrule
	\end{tabular}
	
	\vspace{8pt}
	
	\textbf{(b) Intensity-stratified}\\[4pt]
	\begin{tabular}{l r rrr rrr rrr}
		\toprule
		& & \multicolumn{3}{c}{QiFeng} & \multicolumn{3}{c}{ERA5} & \multicolumn{3}{c}{CCMP} \\
		\cmidrule(lr){3-5}\cmidrule(lr){6-8}\cmidrule(lr){9-11}
		Category & $N$ & MAE & Bias & RMSE & MAE & Bias & RMSE & MAE & Bias & RMSE \\
		\midrule
		TD & 675 & 4.9 & $+4.1$ & 6.4 & 2.4 & $-1.1$ & 2.9 & 2.2 & $-0.6$ & 2.8 \\
		TS & 755 & 5.3 & $+1.8$ & 7.1 & 6.3 & $-6.2$ & 7.3 & 5.5 & $-4.9$ & 6.6 \\
		Cat 1--2 & 228 & 8.0 & $-3.1$ & 9.9 & 18.5 & $-18.5$ & 19.4 & 14.9 & $-14.8$ & 16.7 \\
		Cat 3 & 57 & 10.0 & $-8.8$ & 11.8 & 28.3 & $-28.3$ & 28.8 & 23.1 & $-23.1$ & 24.2 \\
		Cat 4--5 & 64 & 16.3 & $-15.8$ & 18.7 & 39.6 & $-39.6$ & 40.3 & 34.9 & $-34.9$ & 36.1 \\
		\midrule
		All & 1779 & 6.0 & $+1.1$ & 8.2 & 8.3 & $-7.8$ & 12.7 & 7.1 & $-6.2$ & 11.1 \\
		\bottomrule
	\end{tabular}
\end{table}

\subsubsection{Independent Comparison with Dropsondes}
\label{sec:val-dropsonde}

To conduct an independent validation of QiFeng's full-scale reconstruction results, we used point-by-point spatiotemporal matching with dropsonde observation data from 2020--2022. Considering the fast descent process of dropsondes and the temporal evolution of TC structures, the spatiotemporal matching parameters were set to a $\pm 0.5$ hour time window and a 10\,m height tolerance, consistent with the dropsonde validation strategy in \cite{han2025grl}.

\begin{figure}[htbp]
	\centering
	\includegraphics[width=\textwidth]{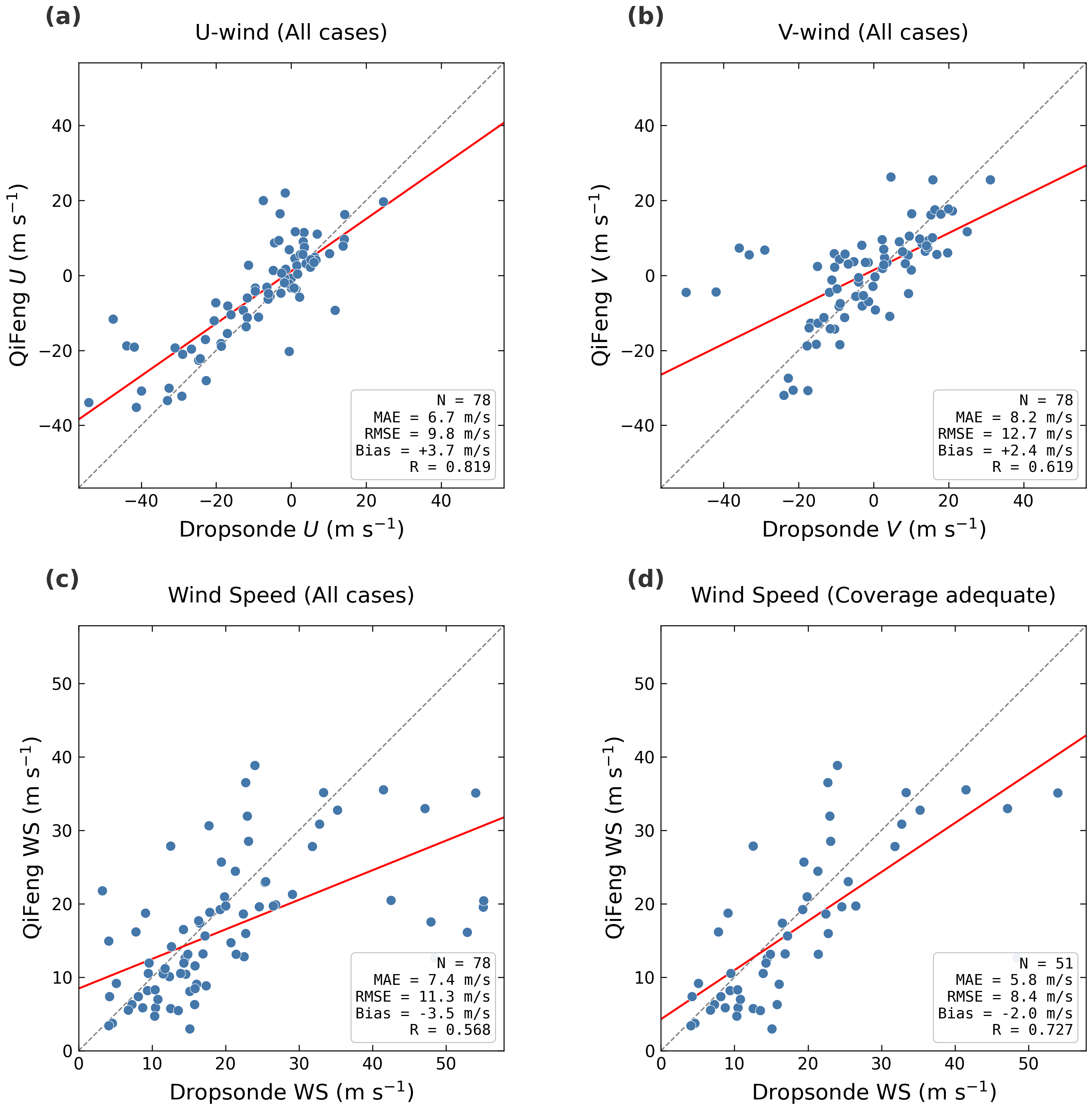}
	\caption{Scatter plot comparison between QiFeng reconstruction and dropsonde observations (2020--2022).
		(a) $U$-wind component, all cases; (b) $V$-wind component, all cases; (c) wind speed, all cases; (d) wind speed, OCS-qualified (coverage adequate) subset.
		Red solid lines are linear regression fits; gray dashed lines are 1:1 reference lines.}
	\label{fig:dropsonde-scatter}
\end{figure}

Fig.~\ref{fig:dropsonde-scatter} displays scatter plots of the fully matched data under different observation criteria. The reconstructed wind fields show good agreement with dropsonde observations in the low to moderate wind speed range, with scatter decreasing as the OCS criterion tightens. However, in the high wind speed regime, the reconstructed wind speeds exhibit underestimation and larger divergence, primarily limited by the CYGNSS L-band signal's sensitivity saturation effect under extreme sea states and interference from heavy precipitation attenuation.

Table~\ref{tab:dropsonde-metrics} summarizes the quantitative error statistics. Across all 24 TC cases (78 paired points), the wind speed RMSE is 11.3\,m\,s$^{-1}$ ($R = 0.568$). After applying the OCS criterion (13 cases, 51 paired points), the RMSE drops to 8.4\,m\,s$^{-1}$ ($R = 0.727$, a 26\% reduction), consistent with the design motivations of the OCS criterion.

\begin{table}[htbp]
	\centering
	\caption{Point-by-point quantitative error statistics between QiFeng reconstruction and matched dropsondes (2020--2022, time window $\pm 0.5$\,h, height tolerance 10\,m).
		``All'' refers to all matched samples; ``OCS'' (Observation Coverage Sufficiency) refers to the subset meeting the OCS criterion ($n_\mathrm{obs} \geq 300$, cells100 $\geq 7$, az\_cov $\geq 0.625$).
		Units: MAE, Bias, RMSE in m\,s$^{-1}$.}
	\label{tab:dropsonde-metrics}
	\setlength{\tabcolsep}{4pt}
	\begin{tabular}{l r r rrrrr rrrrr rrrrr}
		\toprule
		& & & \multicolumn{5}{c}{$U$ (m\,s$^{-1}$)} & \multicolumn{5}{c}{$V$ (m\,s$^{-1}$)} & \multicolumn{5}{c}{Wind Speed (m\,s$^{-1}$)} \\
		\cmidrule(lr){4-8}\cmidrule(lr){9-13}\cmidrule(lr){14-18}
		Subset & Cases & $N$ & MAE & Bias & RMSE & $R$ && MAE & Bias & RMSE & $R$ && MAE & Bias & RMSE & $R$ & \\
		\midrule
		All & 24 & 78 & 6.7 & $+3.7$ & 9.8 & 0.819 && 8.2 & $+2.4$ & 12.7 & 0.619 && 7.4 & $-3.5$ & 11.3 & 0.568 & \\
		OCS & 13 & 51 & 6.5 & $+3.5$ & 9.9 & 0.822 && 6.1 & $-0.5$ & 7.8 & 0.821 && 5.8 & $-2.0$ & 8.4 & 0.727 & \\
		\bottomrule
	\end{tabular}
\end{table}

A spatial wind vector comparison for TC PAULETTE (2020) confirms that QiFeng captures the overall circulation pattern, though local positional offsets appear near the eyewall due to structural evolution within the 0.5-hour matching window (Supplementary Figure~S3).

\subsubsection{Independent Comparison with TDR Observations and Physical Constraint Ablation Experiment}
\label{sec:val-tdr}

We further utilized Tail Doppler Radar (TDR) data for validation. TDR can provide three-dimensional wind field structures within the TC. We selected the 0.5\,km altitude layer---the lowest available altitude layer in TDR merge products---and corrected it to a 10\,m altitude to compare with the QiFeng reconstructions.

The TDR 0.5\,km altitude is located near the peak of the TC boundary layer jet, where wind speeds are systematically higher than the sea surface 10\,m wind speed. To ensure physical consistency, we applied a radially dependent height correction based on the statistical dropsonde profiles of \cite{franklin2003dropsonde}---a larger reduction factor ($f \approx 0.76$) near the eyewall and a smaller one ($f \approx 0.87$) in the outer vortex (Section~\ref{sec:tdr-correction}; Supplementary Section~S9). To obtain a sufficient volume of matched samples, the time window was relaxed to $\pm 1.5$ hours.

\begin{figure}[htbp]
	\centering
	\includegraphics[width=\textwidth]{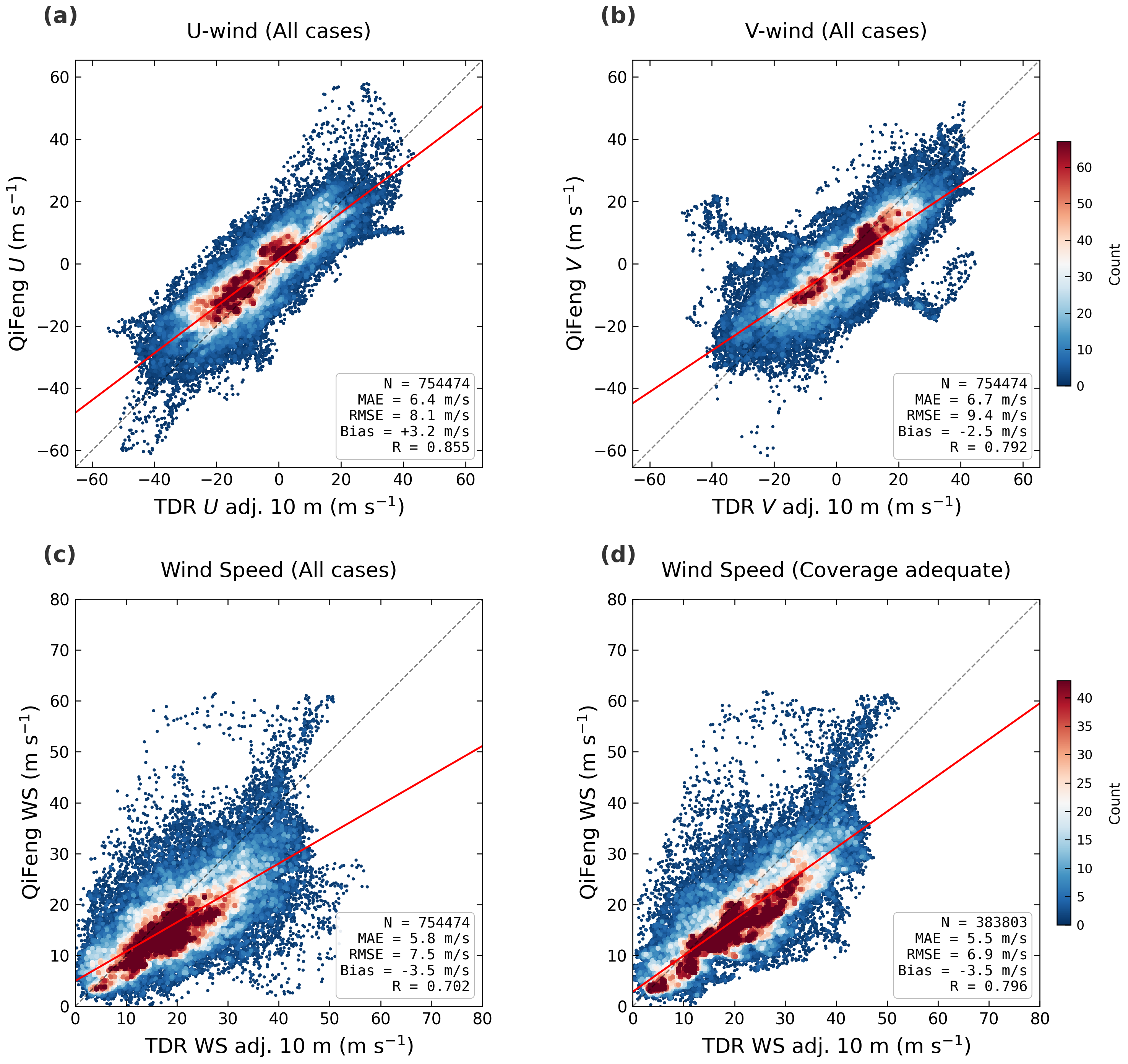}
	\caption{Scatter density plot of QiFeng reconstruction vs.\ TDR corrected 10\,m wind field (2020--2022).
		(a) $U$-wind component, all cases; (b) $V$-wind component, all cases; (c) wind speed, all cases; (d) wind speed, OCS-qualified (coverage adequate) subset.
		TDR 0.5\,km wind speeds were corrected to 10\,m using the radially dependent reduction factor described in Section~\ref{sec:tdr-correction} and Supplementary Section~S9.
		Colors indicate point density; red solid lines are linear regression fits; gray dashed lines are 1:1 reference lines.}
	\label{fig:tdr-scatter}
\end{figure}

Fig.~\ref{fig:tdr-scatter} displays a scatter density plot, where the core high-density region is distributed around the diagonal line. Compared to the dropsonde results, the TDR comparison shows more obvious dispersion, primarily because relaxing the time window to 1.5 hours introduces evolutionary errors of fine TC structures.

Table~\ref{tab:tdr-metrics} summarizes the grid-point-by-grid-point quantitative error statistics, including wind direction fidelity metrics and an ablation without physical constraints ($\lambda=0$). The TDR 201$\times$201 (2\,km resolution) grid, after correction to 10\,m (Section~\ref{sec:tdr-correction}), was nearest-neighbor interpolated to the QiFeng 256$\times$256 (1.5\,km) grid for pixel-by-pixel comparison. A total of 47 cases provided over 750,000 valid paired points.

\textbf{Vector wind field accuracy} ($\lambda=15$). Across all 47 cases (over 750,000 matched points), the wind speed RMSE is 7.5\,m\,s$^{-1}$ ($R = 0.702$) and wind direction RMSE is $35.1^\circ$. After applying the OCS criterion (23 cases), the wind speed RMSE decreases to 6.9\,m\,s$^{-1}$ ($R = 0.796$), and the wind direction RMSE decreases to $28.1^\circ$ (a 20\% reduction), reaffirming the screening effect of the OCS criterion (full component metrics in Table~\ref{tab:tdr-metrics}).

\textbf{Role of physical constraints} ($\lambda=0$ ablation). To verify the effectiveness of physical constraints under real observation conditions, we ran QiFeng reconstructions with $\lambda=0$ (removing all physical constraints) on the same batch of 47 TDR-matched cases. The results (Table~\ref{tab:tdr-metrics}, bottom half) show that without physical constraints, the wind direction RMSE spiked from $35.1^\circ$ to $88.0^\circ$---approaching the theoretical value for random directions ($\sim$104$^\circ$)---and the $V$ component correlation plummeted near zero ($R = -0.024$). Notably, the scalar wind speed RMSE remained almost unchanged (7.5 vs. 7.6\,m\,s$^{-1}$), demonstrating that the core function of the physical constraints is to resolve the scalar-to-vector directional ambiguity rather than improve wind speed magnitude. This matches the conclusion from the OSSE ablation experiment in Section~\ref{sec:lambda-selection} where $\lambda=0$ yielded a wind direction RMSE as high as $73.3^\circ$, and provides further confirmation under real error sources including altitude corrections, time-window matching, and TC center offsets.

\begin{table}[htbp]
	\centering
	\caption{QiFeng vs.\ TDR corrected 10\,m wind field: grid-point error statistics and physical constraint ablation (2020--2022, $\pm 1.5$\,h window).
		TDR corrected to 10\,m via the radially dependent reduction factor (Section~\ref{sec:tdr-correction}).
		Top half: $\lambda=15$ (full constraints); bottom half: $\lambda=0$ (no constraints).
		``All'' = all matched samples; ``OCS'' = OCS-qualified subset (Section~\ref{sec:coverage-criteria}).
		Units: m\,s$^{-1}$ (MAE, Bias, RMSE); $^\circ$ (wind direction). \textbf{Bold} = best; \underline{underline} = second-best.}
	\label{tab:tdr-metrics}
	\setlength{\tabcolsep}{3pt}
	\resizebox{\textwidth}{!}{%
		\begin{tabular}{l l r r rrrrr rrrrr rrrrr rrr}
			\toprule
			& & & & \multicolumn{5}{c}{$U$ (m\,s$^{-1}$)} & \multicolumn{5}{c}{$V$ (m\,s$^{-1}$)} & \multicolumn{5}{c}{Wind Speed (m\,s$^{-1}$)} & \multicolumn{3}{c}{Wind Direction} \\
			\cmidrule(lr){5-9}\cmidrule(lr){10-14}\cmidrule(lr){15-19}\cmidrule(lr){20-22}
			$\lambda$ & Subset & Cases & $N$ & MAE & Bias & RMSE & $R$ && MAE & Bias & RMSE & $R$ && MAE & Bias & RMSE & $R$ && Dir-RMSE ($^\circ$) & $V_t$-RMSE & $V_r$-RMSE \\
			\midrule
			\multirow{2}{*}{15}
			& All & 47 & 754\,474 & \textbf{6.4} & \tb{$+3.2$} & \textbf{8.1} & \textbf{0.855} && \textbf{6.7} & \tb{$-2.5$} & \textbf{9.4} & \textbf{0.792} && \textbf{5.8} & \tb{$-3.5$} & \textbf{7.5} & \textbf{0.702} && \textbf{35.1} & \textbf{9.2} & \textbf{8.4} \\
			& OCS & 23 & 383\,803 & \textbf{6.3} & \tb{$+2.8$} & \textbf{8.1} & \textbf{0.879} && \textbf{6.7} & \tb{$-2.7$} & \textbf{8.6} & \textbf{0.862} && \textbf{5.5} & \tb{$-3.5$} & \textbf{6.9} & \textbf{0.796} && \textbf{28.1} & \textbf{7.9} & \textbf{8.8} \\
			\midrule
			\multirow{2}{*}{0}
			& All & 47 & 754\,474 & 10.1 & $+3.9$ & 16.1 & 0.346 && 13.9 & $-3.5$ & 19.6 & $-0.024$ && 6.0 & $-3.7$ & 7.6 & 0.701 && 88.0 & 23.8 & 9.5 \\
			& OCS & 23 & 383\,803 & 11.3 & $+3.7$ & 18.0 & 0.315 && 15.6 & $-3.9$ & 21.9 & $-0.081$ && 5.7 & $-3.9$ & 7.1 & 0.791 && 92.6 & 26.6 & 10.4 \\
			\bottomrule
	\end{tabular}}
\end{table}

\begin{figure}[htbp]
	\centering
	\includegraphics[width=\textwidth]{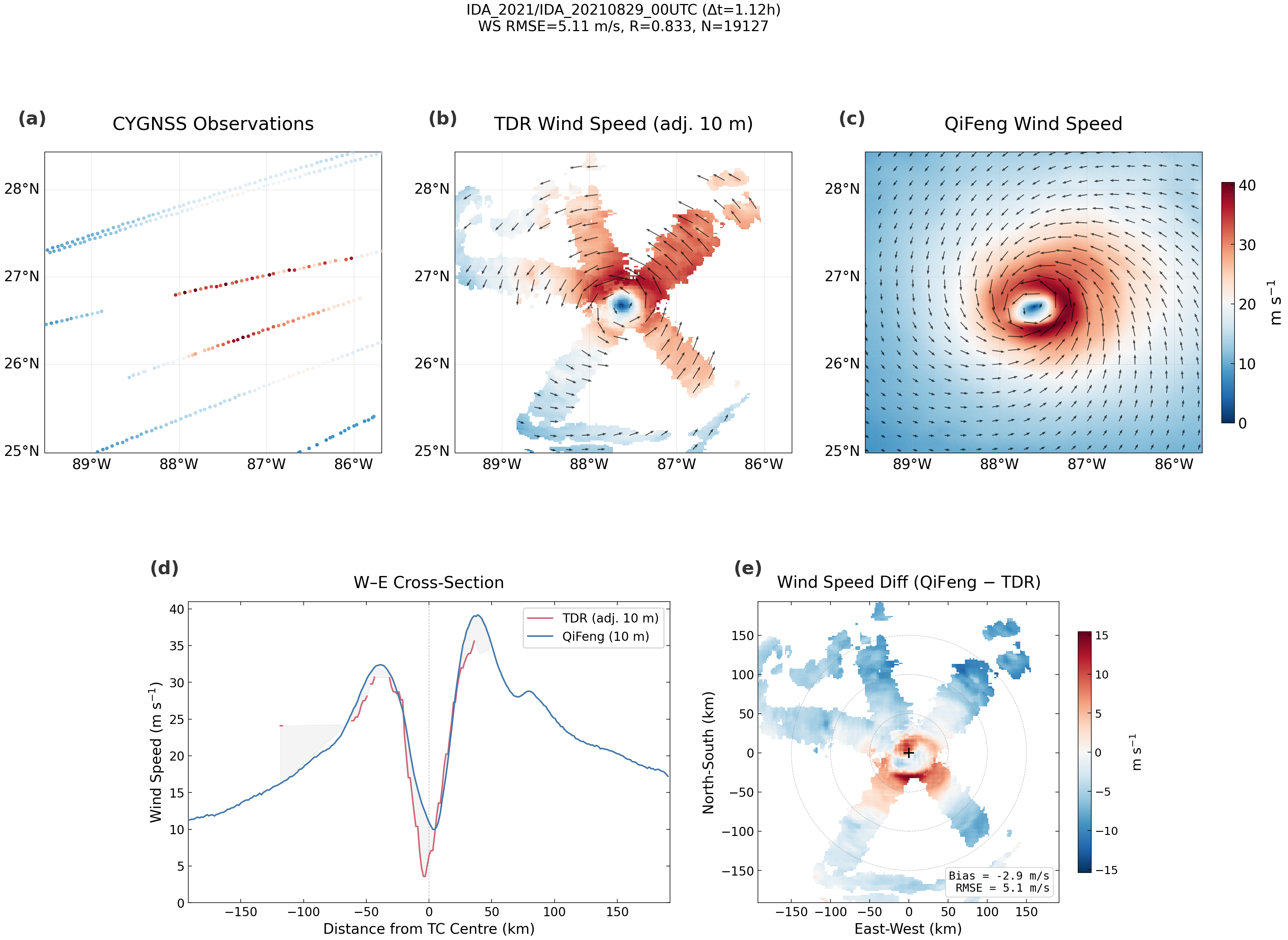}
	\caption{Comparison between QiFeng reconstruction and TDR-corrected 10\,m wind field for TC IDA (2021-08-29 00UTC, $V_{\max} = 90$\,kt, Cat\,2; QiFeng-TDR time offset $\pm 1.12$\,h, within the $\pm 1.5$\,h matching window; 19,127 matched points).
		(a) CYGNSS observation distribution (colors denote wind speed), with multiple tracks covering the eyewall and periphery but not crossing the TC eye;
		(b) TDR-corrected 10\,m wind speed field and wind vectors;
		(c) QiFeng reconstructed wind speed field and wind vectors;
		(d) W-E cross-eye wind speed profile comparison (TDR red, QiFeng blue);
		(e) Wind speed difference field (QiFeng $-$ TDR).}
	\label{fig:tdr-quiver-examples}
\end{figure}

Fig.~\ref{fig:tdr-quiver-examples} uses TC IDA (2021-08-29 00UTC, IBTrACS $V_{\max} = 90$\,kt, Cat\,2) as an example to show the spatial comparison between the QiFeng reconstruction and the TDR-corrected 10\,m wind field. IDA was one of the strongest and most destructive TCs in the Atlantic in 2021, undergoing rapid intensification prior to landfall in Louisiana at this time. Panel (a) shows the CYGNSS observation distribution, with multiple tracks covering the eyewall and outer regions but not directly crossing the TC eye---a coverage pattern similar to the OSSE No-Eye scenario (Scenario B, Section~\ref{sec:osse-design}), providing a typical case to evaluate QiFeng's inner-core reconstruction capability without direct eye observations. Panel (b) shows the TDR-corrected 10\,m wind speed field, displaying a significantly asymmetric eyewall structure, with peak wind speeds ($\sim$35--40\,m\,s$^{-1}$) concentrated in the northeast quadrant. Panel (c) shows QiFeng reconstructed an identifiable TC eye and eyewall structure despite the lack of direct eye observations, and the eyewall's asymmetric distribution---with the northeast quadrant stronger than the southwest---qualitatively matched the TDR observations. In the W-E cross-eye profile (Panel d), QiFeng (blue) roughly tracked TDR's (red) bimodal eyewall structure in the inner core region ($|r| < 80$\,km), with the radial positions of the peaks and eye radius largely consistent with TDR.

Quantitatively, this case achieved a wind speed RMSE of 5.1\,m\,s$^{-1}$, $R = 0.833$, and $U$/$V$ component correlations of 0.919 and 0.902 across 19,127 valid matched grid points (27.2\% coverage), indicating strong agreement with independent TDR observations in both scalar wind speed and vector directional structure. However, the wind speed difference field (Fig.~\ref{fig:tdr-quiver-examples}e) reveals spatially structured errors: the eyewall region ($|r| \approx 30$--$60$\,km) exhibits an alternating positive-negative error pattern reflecting minor positional shifts of fine eyewall structures, while the outer region ($|r| > 100$\,km) exhibits a systematic negative bias ($-2.9$\,m\,s$^{-1}$). This outer-region underestimation arises from two compounding factors: CYGNSS observations at the domain edges predominantly captured low wind speed signals, biasing the likelihood guidance downward, while the diffusion prior defaults to climatological mean wind speeds in sparsely observed areas---both reinforcing low biases in the TC periphery. This spatial error pattern is consistent with the design rationale of the OCS criterion (Section~\ref{sec:coverage-criteria}).

Conversely, TC TEDDY (Cat\,4, 120\,kt) illustrates a failure mode: with only 65 CYGNSS points along a single track beyond 100\,km from the TC center, QiFeng produced a banded flow pattern rather than a closed vortex (RMSE $= 18.0$\,m\,s$^{-1}$, the worst among all 47 TDR cases; Supplementary Figure~S5). This confirms that a single track lacking angular diversity cannot constrain a 2D vector vortex, directly motivating the OCS criterion (Section~\ref{sec:coverage-criteria}).

\subsubsection{Multi-source Observation Fusion Case Study: TC FIONA and TDR Comparison}
\label{sec:fusion-case-study}

The dropsonde and TDR validations above demonstrated that QiFeng can reconstruct reasonable TC vector wind fields from sparse CYGNSS scalar observations alone, but inner-core wind speeds still exhibit underestimation in cases with extremely limited observation coverage. QiFeng's likelihood guidance framework inherently supports the joint assimilation of heterogeneous multi-source observations---requiring only the addition of new observation points, their corresponding observation operators, and error models into the likelihood term---providing a direct pathway to alleviate wind speed underestimation under insufficient CYGNSS coverage by introducing a small number of vector observations. As gold standard in situ observations for TC wind fields, dropsondes directly provide vector winds at 10\,m height ($u$, $v$ components). Their observation operator is a linear identity mapping $H_\text{drop}(\mathbf{x})_k = (u_{i_k,j_k},\, v_{i_k,j_k})$, which inherently complements CYGNSS's nonlinear scalar operator $H(\mathbf{x})=\sqrt{u^2+v^2}$: CYGNSS provides a wide-area wind speed constraint, while dropsondes offer precise wind direction anchoring at discrete points. This section takes TC FIONA (2022-09-18 12UTC) as an example to demonstrate the improvements from jointly assimilating CYGNSS and dropsonde data, using concurrent TDR composite wind fields as an independent validation benchmark.

\begin{figure}[htbp]
	\centering
	\includegraphics[width=\textwidth]{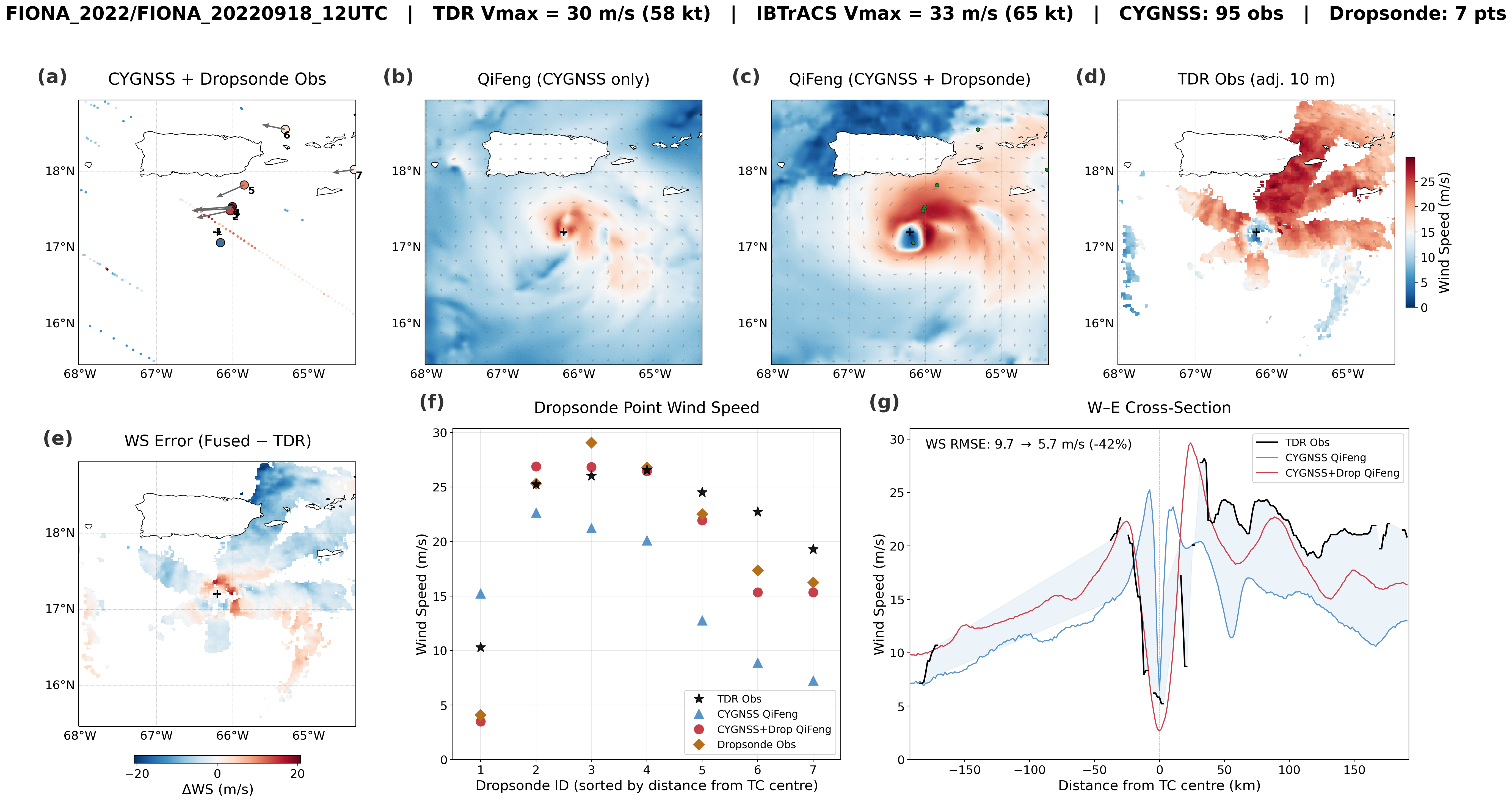}
	\caption{Multi-source observation fusion case study for TC FIONA (2022-09-18 12UTC, IBTrACS $V_{\max} = 33$\,m\,s$^{-1}$, 65\,kt).
		(a) Observation distribution: CYGNSS specular points (95 pts, blue dots) and dropsonde locations (11 pts, numbered circles); red dashed line is the TC track;
		(b) QiFeng reconstruction using only CYGNSS observations;
		(c) QiFeng reconstruction jointly assimilating CYGNSS + dropsondes;
		(d) TDR observed wind speed field (corrected to 10\,m per Section~\ref{sec:tdr-correction});
		(e) Wind speed difference field (fusion result $-$ TDR);
		(f) Point-by-point wind speed comparison at dropsonde locations (sorted by distance to TC center): TDR (black stars), dropsonde (orange diamonds), CYGNSS-only (blue triangles), CYGNSS+Drop (red circles);
		(g) W-E cross-eye wind speed profile: TDR (black), CYGNSS-only (blue, shading = ensemble spread), CYGNSS+Drop (red).}
	\label{fig:fusion-fiona}
\end{figure}

Fig.~\ref{fig:fusion-fiona} displays the multi-source fusion comparison for TC FIONA (2022-09-18 12UTC, IBTrACS $V_{\max} = 33$\,m\,s$^{-1}$, 65\,kt). In this case, CYGNSS coverage is extremely sparse, with only 95 observation points (Panel a) distributed primarily in the TC's outer region and almost no inner-core coverage---far below the $n_\mathrm{obs} \geq 300$ threshold of the OCS criterion in Section~\ref{sec:coverage-criteria}. Concurrently, aircraft reconnaissance missions deployed 11 dropsondes (numbered 1--11 in Panel a) at this time, partially covering the eyewall and inner-core regions. TDR observations (Panel d, corrected to 10\,m per Section~\ref{sec:tdr-correction}) show that FIONA exhibited a significantly asymmetric eyewall structure: peak wind speeds concentrated in the southeast quadrant, reaching about 25--28\,m\,s$^{-1}$ (TDR $V_{\max} = 30$\,m\,s$^{-1}$), while the northwest quadrant was markedly weaker, with eye wind speeds dropping below 5\,m\,s$^{-1}$.

Panel (b) presents the QiFeng reconstruction results using only CYGNSS observations. Due to the severe insufficiency of 95 observation points, the CYGNSS-only reconstruction exhibits a relatively smooth wind field distribution, failing to clearly depict the fine structures of the TC eye and eyewall, resulting in a systemic underestimation of overall wind speed---peak wind speeds in the eyewall region are only about 15--18\,m\,s$^{-1}$, less than 60\% of the TDR observed values. The asymmetric features of the wind field were also not effectively captured, as the diffusion prior tended to generate symmetric structures close to climatological averages in the absence of sufficient observational constraints. According to the OCS criterion in Section~\ref{sec:coverage-criteria}, CYGNSS coverage in this case is far from adequate, representing a scenario where independent reconstruction is not recommended. Panel (c) shows the results of joint assimilation of CYGNSS and dropsondes. After introducing the vector wind observations from the 11 dropsondes, the reconstruction quality improved noticeably: the TC eye became identifiable, eyewall wind speed gradients steepened, and the asymmetric structure (strong winds in the southeast quadrant, relatively weaker in the northwest) qualitatively matched the TDR observations (Panel d). The vector information from the dropsondes anchored the wind directions at discrete points, breaking the inherent directional ambiguity in purely scalar observations, and allowing physical constraints to more effectively propagate directional information from observation points to the surrounding areas. Panel (e), the wind speed error field (fusion result $-$ TDR), shows that the error magnitude in the inner-core region is primarily controlled within $\pm$5\,m\,s$^{-1}$, with local alternating positive and negative error patterns near the eyewall, reflecting minor positional shifts in fine structures. Panel (f) directly compares four sources of wind speed at the 11 dropsonde deployment locations (sorted by distance to the TC center): TDR observations (black stars), dropsonde observations (orange diamonds), CYGNSS-only QiFeng (blue triangles), and CYGNSS+Dropsonde QiFeng (red circles). The CYGNSS-only reconstruction systematically underestimates wind speed compared to TDR and dropsonde observations at almost all locations, with biases being particularly significant in the high-wind inner-core region---for example, at dropsonde locations 3--6 (near the eyewall), TDR and dropsonde observed wind speeds of about 25--29\,m\,s$^{-1}$, while CYGNSS-only reconstructed only 13--20\,m\,s$^{-1}$, an underestimation of 30\%--50\%. After fusing dropsondes (red circles), the point-by-point wind speeds shifted significantly toward TDR and dropsonde observations, shrinking the bias to within 3\,m\,s$^{-1}$ at most locations, improving the underestimation problem in the inner-core high wind speed area. Panel (g) is the West-East (W-E) cross-eye wind speed profile comparison along the TC center. The TDR observation (black solid line) exhibits a typical asymmetric bimodal structure, with the east eyewall peak around 25\,m\,s$^{-1}$, the west peak around 22\,m\,s$^{-1}$, and eye wind speed plunging below 5\,m\,s$^{-1}$. The CYGNSS-only QiFeng (blue solid line, with the light blue shaded area representing ensemble spread) significantly underestimates across the entire profile, failing to track the TDR bimodal structure, with shifted peak positions, insufficient amplitude, and large ensemble spread ($\sim$5\,m\,s$^{-1}$), reflecting high uncertainty when reconstructing under sparse observational conditions. The CYGNSS+Dropsonde fusion result (red solid line) significantly improved: the overall profile shape more closely approximates the TDR observation, the radial positions and relative intensities of the bimodal peaks are better reconstructed, and the wind speed drop in the eye is clearer. Quantitatively, the wind speed RMSE of the cross-eye profile dropped drastically from 9.7\,m\,s$^{-1}$ for CYGNSS-only to 5.7\,m\,s$^{-1}$ for the fusion result, a 42\% reduction. This case demonstrates that even under extremely sparse conditions where CYGNSS coverage is far from sufficient, introducing a small number of dropsonde vector observations can bring noticeable improvements in reconstruction quality within the QiFeng framework (a 42\% reduction in cross-eye profile RMSE). The physical mechanism behind this result lies in the complementary information dimensions of CYGNSS scalar observations and dropsonde vector observations: CYGNSS's wide-area coverage provides global wind speed constraints for the diffusion prior, while dropsonde vector information breaks the scalar-to-vector directional degeneracy at key locations, concentrating QiFeng's conditional posterior distribution from a broad multimodal space to a focused single mode. This also provides methodological validation for future operational applications---using QiFeng as a real-time multi-source observation fusion platform to jointly assimilate heterogeneous observation sources such as CYGNSS, dropsondes, and SFMR.

Ensemble uncertainty estimation via 16 independent diffusion samplings for TC IAN (14 retained after a cyclonic-rotation post-filter that removes members with anti-cyclonic inner-core flow) shows that the spread is radially dependent: largest in the eyewall ($\sim$6--10\,m\,s$^{-1}$) and small ($<$3\,m\,s$^{-1}$) in the eye and outer regions, consistent with the spatial distribution of wind speed gradients and directional ambiguity (Supplementary Figure~S4).

\subsubsection{Physical Constraint Weight \texorpdfstring{$\lambda$}{lambda} Selection}
\label{sec:lambda-selection}

All experiments use a physical constraint weight $\lambda=15$, selected based on a systematic OSSE scan over $\lambda \in \{0, 0.1, \dots, 25\}$ ($n=366$ test samples, Scenario B). As $\lambda$ increases from 0 to 15, the wind direction RMSE drops from $73.3^\circ$ to $18.9^\circ$ (a 74\% reduction) while $V_{\max}$ accuracy remains stable; beyond $\lambda=15$, marginal improvements saturate (Supplementary Table~S2).



\section{Discussion}
\label{sec:discussion}

QiFeng---a physics-guided score-based diffusion framework---reconstructs 1.5\,km resolution TC 2D vector wind fields $(u_{10}, v_{10})$ from sparse, directionless CYGNSS scalar observations by fusing a nonlinear likelihood constraint with three TC boundary-layer physical regularisations during reverse diffusion, resolving the directional ambiguity inherent in scalar-to-vector reconstruction and demonstrating reliable performance in both OSSEs and large-scale real-data experiments.

\paragraph{Extending assimilation paradigms from linear to nonlinear observation operators.}
Existing diffusion-based data assimilation---DiffDA \cite{huang2024diffda} and score-based DA \cite{rozet2023sda}---primarily handles linear observation operators with one-to-one or low-rank state-to-observation mappings. QiFeng's core methodological contribution is generalising SDA to a nonlinear, many-to-one operator $H(u,v)=\sqrt{u^2+v^2}$: for any observed wind speed, the $(u,v)$ solutions form a circle, making direction unidentifiable from the likelihood alone. We address this through three complementary designs: (1)~observation-wise effective variances (Eq.~\eqref{eq:veff}); (2)~a Huber robust likelihood loss (Eq.~\eqref{eq:obs-loss}); and (3)~three physical constraints (divergence, inflow angle, vorticity) injecting directional prior knowledge. The $\lambda$-sensitivity scan (Supplementary Table~S2) and the TDR ablation (Table~\ref{tab:tdr-metrics}) consistently show that removing constraints ($\lambda=0$) drives wind-direction RMSE to near-random levels ($\sim$73$^\circ$--88$^\circ$) while leaving scalar wind speed RMSE almost unchanged---the constraints resolve directional ambiguity without affecting speed magnitude. The paradigm extends in principle to other GNSS-R retrievals that share this nonlinear structure (for example, sea-surface height inversion from delay-Doppler maps, where the topography-to-observable mapping is similarly many-to-one) and, more broadly, to nonlinear geophysical inverse problems---though each extension warrants dedicated validation.

\paragraph{Capabilities in characterizing TC inner-core structures.}
QiFeng materially advances inner-core characterisation over $\sim$0.25$^\circ$ products: a $V_{\max}$ bias of $+1.5$\,m\,s$^{-1}$ shrinks the systematic underestimation by $\sim$79\% / $\sim$75\% versus ERA5 / CCMP (Table~\ref{tab:struct-errors}), and an RMW bias of $-7.1$\,km contrasts with the $\sim$+50\,km positive bias of $0.25^\circ$ products. The SAR cases (Figs.~\ref{fig:sar-ian}, \ref{fig:sar-hinnamnor}) show this concretely---resolved eye, asymmetric eyewall, steep radial gradients. The improvement has three physical roots: a 1.5\,km grid that resolves fine eyewall structures within typical RMW (30--80\,km); CYGNSS L-band penetration of heavy precipitation, where traditional scatterometers fail; and a diffusion prior that learns the statistical diversity of TC structures from high-resolution HWRF simulations and provides physically reasonable infilling in sparsely observed regions.

\paragraph{Implications for global TC monitoring and basins lacking routine aircraft reconnaissance.}
Of $\sim$80--90 TCs annually, only North Atlantic and Eastern Pacific storms ($\sim$25\% of global TC activity) have routine aircraft reconnaissance, while the most active basin---the Western North Pacific ($\sim$30\% of the global total)---along with the Northern Indian Ocean and the Southern Hemisphere relies almost entirely on satellite observations with only sporadic non-routine flights \cite{aberson2006thirtyyears}. Our 249-TC dataset spans all six major basins (Fig.~\ref{fig:framework}d), with the Southern Hemisphere (SI + SP) contributing 35.6\% of snapshots, and basin-stratified statistics show performance in non-reconnaissance basins comparable to the North Atlantic (WP $R=0.845$, SI $R=0.736$, SP $R=0.758$; Table~\ref{tab:basin-intensity}); the Southern Hemisphere TC EMNATI case (Supplementary Figure~S1) confirms that the coordinate-flipping strategy generalises to clockwise circulations. Although HWRF/HAFS produces high-resolution analyses for all global TCs, their inner-core structures in basins without routine reconnaissance are primarily model-driven (Section~\ref{sec:introduction}); QiFeng provides kilometre-scale vector winds anchored directly to satellite surface-wind observations (CYGNSS), offering an independent observational reference for TC intensity calibration, structural analysis and cross-validation of NWP analyses in those basins.

\paragraph{Accuracy and physical consistency of wind direction reconstruction.}
QiFeng's median wind-direction RMSE of $\sim$14$^\circ$ versus ERA5/CCMP (Supplementary Table~S1) lies within the natural variability of TC boundary-layer inflow angles ($15^\circ$--$30^\circ$, \cite{kepert2001tcbl}), and the OCS-filtered TDR validation gives $28.1^\circ$ (Table~\ref{tab:tdr-metrics}); the residual is dominated by fine-structure evolution within the $\pm 1.5$\,h matching window rather than systematic directional bias---consistent with a near-zero directional bias of $-2^\circ$ to $-3^\circ$. Tangential wind RMSE exceeds radial wind RMSE simply because $V_t$ is larger ($\sim$30--60\,m\,s$^{-1}$) than $V_r$ ($\sim$5--15\,m\,s$^{-1}$); in relative terms $V_r$ is in fact more uncertain because it is inferred entirely from the physical constraints without direct observational support, which is why the inflow-angle constraint is decisive---it directly governs the $V_r$/$V_t$ allocation.

\paragraph{OCS and operational application potential.}
The OCS criterion retains $\sim$40\% of the $V_{\max}$-validated samples (1{,}779 of 4{,}450) at $V_{\max}$ $R=0.795$, surpassing the full-sample baselines of ERA5 (0.754) and CCMP (0.756). Because OCS depends only on CYGNSS observation metrics without external references, it directly supports real-time operational screening, and the dropsonde (Table~\ref{tab:dropsonde-metrics}) and TDR (Table~\ref{tab:tdr-metrics}) independent validations confirm that the OCS subset uniformly outperforms the full sample. As the GNSS-R constellation expands (FY-3E GNOS-II, Tianmu-1, etc.), the OCS-passing fraction is expected to grow substantially.

\paragraph{Near real-time application feasibility.}
Although our experiments use retrospective IBTrACS and CYGNSS CDR v3.2 scientific-grade products, QiFeng is computationally light enough for near real-time (NRT) deployment: a $\pm$3\,h snapshot reconstructs in $\sim$197\,s ($\sim$3.3\,min) on a single NVIDIA RTX\,3090. CYGNSS NRT products carry a 1--2\,h delay, so end-to-end latency from overpass to vector wind field would be dominated by the CYGNSS delay itself, with QiFeng adding only $\sim$3\,min on top---giving $\sim$1--2\,h overall, compared with $\sim$5\,h for ERA5T (preliminary), $\sim$5 days for ERA5 (final) and $>$1\,day for CCMP. Operationalisation still requires (i)~replacing IBTrACS centring with real-time agency advisories (e.g., NHC / JTWC), (ii)~validating NRT data quality, and (iii)~scheduling multi-TC parallel processing. Diffusion-acceleration techniques such as consistency models \cite{song2023consistency} and progressive distillation \cite{salimans2022progressive} are further expected to reduce the current 64-step sampler to 1--4 steps, a 1--2 order-of-magnitude speed-up.

\paragraph{Information complementarity in multi-source fusion.}
The TC FIONA fusion case demonstrates information complementarity within QiFeng's likelihood-guidance framework: with only 95 CYGNSS points (far below the 300 OCS threshold), adding 11 dropsondes cuts the cross-eye profile RMSE by 42\%. CYGNSS provides a wide-area scalar wind-speed constraint, dropsondes anchor direction at discrete locations, and their synergy collapses the conditional posterior from a broad multimodal state to a concentrated single mode. The same machinery natively accommodates SFMR along-track wind speeds, SAR area wind speeds and scatterometer vector winds in future operational analyses.

\paragraph{Limitations.}
Several limitations remain. (1)~Wind speeds in extreme regimes ($V_{\max}>60$\,m\,s$^{-1}$) may be underestimated, constrained by L-band sensitivity saturation at high wind speeds and the limited Cat\,5 sample in the HWRF training data. (2)~Independent point-by-point validation (dropsonde, TDR) is concentrated in the North Atlantic because of where reconnaissance flies; although the diffusion prior and physical-constraint designs are basin-agnostic (HWRF training spans multiple basins; NICAM pre-training covers all major basins over 30 years), and basin-stratified statistics (Table~\ref{tab:basin-intensity}) plus the Southern Hemisphere SAR case (Supplementary Figure~S1) indirectly support cross-basin consistency, point-wise quantitative validation in the Western North Pacific, Northern Indian Ocean, and Southern Hemisphere awaits future multi-source data (expanded reconnaissance, new buoy networks). (3)~The diffusion prior is trained on HWRF simulations---the same model whose inner-core analyses we argue are primarily model-driven---but this is not circular: the prior learns the \emph{statistical distribution} of TC wind-field structures (asymmetries, eyewall gradients, radial-tangential partitioning) from $\sim$1{,}130 HWRF snapshots rather than reproducing any specific analysis, and at inference CYGNSS observations select the realisation consistent with the actual measurements; nevertheless QiFeng may inherit HWRF structural biases (e.g., insufficient extreme asymmetries) and should be regarded as complementary to, not a replacement for, operational analyses. (4)~The 2020--2022 sample distribution of strong (Cat\,4--5) TCs reflects natural variability; long-term climatological evaluation is left to future work. (5)~QiFeng's applicability is most fundamentally constrained by CYGNSS coverage itself: of 10{,}067 TC reporting times in 2018--2022, 32.8\% have no CYGNSS observations within the $\pm$3\,h window and only 24.9\% of those with observations meet OCS (Supplementary Table~S3), so for the majority of TC snapshots---including rapidly evolving stages where structural information matters most---no reliable reconstruction is produced; under sparse-but-present coverage, the prior dominates and outputs may skew toward climatological mean structures rather than individual realisations. (6)~Domain definition relies on externally provided TC centre positions; this paper uses IBTrACS best-tracks, whose retrospective accuracy is high but delay is multi-month, while NRT advisories (NHC/JTWC, ATCF b-decks) have hour-scale delay comparable to CYGNSS NRT but inherent $\mathcal{O}(10)$\,km uncertainty especially in basins without routine reconnaissance, with centring errors propagating to domain offsets and degrading RMW / eye-structure extraction. The mean reconstructed eye offset of 14.5\,km (median 5.4\,km) indicates tolerance to moderate centring errors, but large localisation errors will still degrade reconstruction quality. (7)~All experiments are retrospective (IBTrACS, CYGNSS CDR v3.2); QiFeng's end-to-end performance on an NRT data stream---and the specific impact of NRT data-quality differences and operational localisation errors---awaits dedicated future assessment.

\paragraph{Future Outlook.}
QiFeng's methodological framework and experimental results lay the groundwork for the following directions: (1) Global GNSS-R constellation fusion: expanding QiFeng to fuse multi-constellation GNSS-R observations such as CYGNSS, FY-3E GNOS-II, and Tianmu-1 can significantly enhance spatiotemporal coverage density. (2) Temporal evolution modeling: currently, QiFeng processes each $\pm$3\,h snapshot independently; introducing diffusion priors with a temporal dimension or conditional generative frameworks can achieve temporally coherent reconstructions of TC wind fields. (3) Operational validation and integration: integrating QiFeng reconstructed wind fields into the initialization processes of storm surge models, wave models, or regional AI typhoon prediction systems such as HITS \cite{niu2026hits} to evaluate their practical improvements in downstream applications. (4) Multi-layer 3D expansion: the current single-level reconstruction cannot constrain the vertical structure of the TC vortex. Key features such as eyewall tilt, the warm-core anomaly, and the boundary-layer-to-outflow transition require multi-level wind and thermodynamic fields that a 10\,m-only product cannot provide. Extending to multi-level joint reconstruction---where surface observations propagate structural information to upper levels through a jointly trained prior that encodes physically consistent vertical covariances---would enable 3D vortex characterization for basins without routine aircraft reconnaissance, addressing the coverage gap identified in the limitations above.

In summary, by combining a score-based diffusion model with a nonlinear observation operator and TC boundary layer physical constraints, QiFeng demonstrates the feasibility of reconstructing high-resolution TC vector wind fields from sparse, directionless GNSS-R scalar observations. The core idea of this framework---utilizing physical constraints to resolve identifiability degradation in nonlinear inverse problems---holds methodological significance extending beyond TC wind field reconstruction, offering a new paradigm for widely existing nonlinear, ill-posed inverse problems in geophysics.


\section{Methods}
\label{sec:method}

QiFeng consists of three modules (Fig.~\ref{fig:framework}): (1) \textbf{a score model trained based on EDM} \cite{karras2022edm}, encoding a physically reasonable TC $(u_{10},v_{10})$ vector wind field prior; (2) \textbf{nonlinear likelihood guidance}, generalizing standard SDA \cite{rozet2023sda} to the nonlinear observation operator $H(\mathbf{x})=\sqrt{u^2+v^2}$; (3) \textbf{three TC boundary layer physical constraints}, injected as gradient guidance into the reverse diffusion process. The following sections focus on detailing our specific designs in nonlinear likelihood guidance, physical constraints, and observation error modeling. Detailed contents of the underlying EDM training framework and the SDA assimilation paradigm can be found in \cite{karras2022edm} and \cite{rozet2023sda}, respectively.

\subsection{Problem Definition}
\label{sec:problem}

Let $\mathbf{x} = (u_{10},\, v_{10}) \in \mathbb{R}^{2\times 256\times 256}$ be the TC 10\,m vector wind field to be reconstructed (1.5\,km resolution, domain 384\,km $\times$ 384\,km). CYGNSS provides scalar wind speed observations at specular points $\{(i_k, j_k)\}_{k=1}^K$ (Eq.~\eqref{eq:obs-model}):
\begin{equation}
	y_k = \sqrt{u_{i_k,j_k}^2 + v_{i_k,j_k}^2} + \varepsilon_k,
	\qquad \varepsilon_k \sim \mathcal{N}(0,\,\sigma_k^2),\quad k=1,\dots,K.
	\label{eq:obs-model}
\end{equation}
The corresponding nonlinear observation operator (Eq.~\eqref{eq:obs-operator}) is:
\begin{equation}
	H(\mathbf{x})_k = \sqrt{u_{i_k,j_k}^2 + v_{i_k,j_k}^2}.
	\label{eq:obs-operator}
\end{equation}
This operator maps a 2D vector state $(u,v)$ to a 1D scalar wind speed and is \emph{nonlinear}: for any observed $y_k$, infinitely many $(u,v)$ combinations satisfy $\sqrt{u^2+v^2}=y_k$, making wind direction highly unidentifiable. This constitutes a fundamental difference from linear observation operators in standard data assimilation, and is the core challenge that physical constraints are needed to help resolve in this study.

\subsection{Diffusion Prior}
\label{sec:edm-prior}

We adopted the EDM framework \cite{karras2022edm} to train a denoising network $D_{\boldsymbol\theta}(\mathbf{x};\sigma)$ (with a DDPM++ SongUNet \cite{song2021sde} backbone), retaining its original designs for noise scheduling, preconditioning, and weighted denoising loss. Here we emphasize two customized designs made specifically for TC $(u,v)$ vector wind fields.

\paragraph{Wind speed consistency auxiliary loss.}
The standard EDM loss computes the $L_2$ error independently per channel, which cannot guarantee consistency between the $u$ and $v$ channels at the synthetic wind speed level. Since the core inverse problem of QiFeng---reconstructing a vector field from scalar wind speeds---is achieved by matching $\sqrt{u^2+v^2}$ with observations, if the denoiser only pursues channel-wise accuracy without focusing on synthetic wind speed, it will introduce a systematic wind speed reconstruction bias during inference. Therefore, we added an auxiliary wind speed consistency loss (Eq.~\eqref{eq:ws-loss}) on top of the standard EDM denoising loss:
\begin{equation}
	\mathcal{L}_\text{ws} = \mathbb{E}\!\left[\lambda(\sigma)\,\Bigl(\sqrt{\hat{u}^2+\hat{v}^2}-\sqrt{u^2+v^2}\Bigr)^{\!2}\right],
	\label{eq:ws-loss}
\end{equation}
where $\hat{u},\hat{v}$ are the denoiser outputs, $u,v$ are the ground truths, and $\lambda(\sigma)$ is the EDM's noise-level-dependent weight. This loss directly constrains the denoiser to maintain the accuracy of the synthetic wind speed $\sqrt{u^2+v^2}$ while reconstructing the $(u,v)$ components, ensuring that the prior model remains consistent at the objective function level with the nonlinear observation operator $H(\mathbf{x})=\sqrt{u^2+v^2}$ in the inference stage.

\paragraph{Two-stage pre-training and fine-tuning strategy.}
High-resolution TC wind field data is scarce (HWRF inner-core domain data volume is limited, detailed in Section~\ref{sec:data}). Directly training a diffusion model is prone to overfitting and failing to capture the statistical diversity of multi-scale TC structures. We designed a cross-resolution two-stage training strategy: the first stage pre-trains on $\sim$55,000 TC snapshots from the NICAM global cloud-resolving simulation \cite{matsuoka2023tcdata}---NICAM's original resolution is about 14\,km, projected, cropped, and interpolated to the $256\times 256$ grid consistent with HWRF (detailed in Section~\ref{sec:data}), supplemented with multi-scale random cropping augmentation. This enables the model to learn the universal statistical structures of TC circulations (asymmetries, spiral rainbands, multi-scale vortices). The second stage transfers the pre-trained weights to the HWRF inner-core domain (1.5\,km) for fine-tuning, allowing the model to adapt to fine eyewall structures and extreme wind speed gradients at high resolution. The training and test sets are strictly split by TC name, ensuring no data leakage (detailed in Section~\ref{sec:data}). Upon completion of training, the score model provides the prior score evaluation $\nabla_{\mathbf{x}}\log p_\sigma(\mathbf{x})\approx(D_{\boldsymbol\theta}(\mathbf{x};\sigma)-\mathbf{x})/\sigma^2$.

\subsection{Nonlinear Likelihood Guidance}
\label{sec:nonlinear-likelihood-guidance}

SDA \cite{rozet2023sda} fuses the diffusion model's prior score $\nabla_{\mathbf{x}}\log p(\mathbf{x})$ with the observation likelihood score $\nabla_{\mathbf{x}}\log p(\mathbf{y}|\mathbf{x})$ during the reverse sampling process. However, its original framework only handles linear observation operators. QiFeng extends this to the nonlinear observation operator $H(\mathbf{x})=\sqrt{u^2+v^2}$. The four key designs in our nonlinear likelihood guidance are detailed below.

\subsubsection{Observation-wise adaptive effective variance}
\label{sec:adaptive-variance}

Standard SDA employs a globally uniform effective variance $V=\sigma_y^2+\sigma_t^2\gamma$ ($\sigma_y$ is the baseline observation noise, $\gamma$ is the approximation error weight), meaning all observations contribute equally to the gradient field. However, the error of CYGNSS YSLF wind speed retrievals is significantly wind-speed-dependent: it is about 2\,m\,s$^{-1}$ below 30\,m\,s$^{-1}$, but rapidly increases to 8--13\,m\,s$^{-1}$ in high-wind regimes due to the sensitivity saturation of L-band GNSS-R signals \cite{ruf2019cygnss, ruf2024characterization} (detailed in Section~\ref{sec:obs-error}). If a uniform variance is still used, high-error observations will produce an overly strong guidance signal in the gradient field, causing the reconstructed wind field to skew toward these unreliable high-wind observations.

Therefore, QiFeng introduces an observation-wise effective variance:
\begin{equation}
	V_{\text{eff},k} = \underbrace{\sigma_y^2 + \sigma_t^2\gamma\vphantom{(\sigma_k/S)^2}}_{\text{Global (SDA framework)}} + \underbrace{(\sigma_k/S)^2}_{\text{Observation-wise error}},
	\label{eq:veff}
\end{equation}
where $\sigma_k$ is the wind-speed-dependent error standard deviation of the $k$-th observation (Section~\ref{sec:obs-error}), and $S=150$\,m\,s$^{-1}$ is the normalization factor. At high wind speeds, $\sigma_k$ is larger, corresponding to a larger $V_{\text{eff},k}$, and the observation's contribution to the gradient is automatically reduced---achieving data-driven, observation-wise adaptive down-weighting without any manual intervention.

\subsubsection{Huber robust likelihood loss}
\label{sec:likelihood}

Real CYGNSS observations inevitably contain outliers caused by rain contamination and anomalous sea states. The standard Gaussian negative log-likelihood ($L_2$ loss) is extremely sensitive to outliers: a single anomalous observation can generate an immense guiding force in the gradient field, severely distorting local wind field reconstructions. To enhance robustness, QiFeng replaces the $L_2$ loss with the Huber function:
\begin{equation}
	\mathcal{L}_\text{obs} = \sum_{k=1}^{K}\frac{1}{V_{\text{eff},k}}\,\text{Huber}_{\delta_k}\!\Bigl(H(\hat{\mathbf{x}}_0)_k - y_k/S\Bigr),
	\quad \delta_k = 3\sqrt{V_{\text{eff},k}},
	\label{eq:obs-loss}
\end{equation}
where $\hat{\mathbf{x}}_0 = D_{\boldsymbol\theta}(\mathbf{x}_t;\sigma_t)$ is the current denoising estimate. The Huber threshold is set to $\delta_k=3\sqrt{V_{\text{eff},k}}$ (the $3\sigma$ rule): residuals within $3\sigma$ are penalized quadratically, consistent with a standard Gaussian likelihood; beyond $3\sigma$, it switches to a linear penalty ($L_1$), effectively limiting the gradient contribution of outlier observations. Note that $\delta_k$ is also observation-wise adaptive, coupled with the effective variance: high-uncertainty observations have a more lenient Huber transition threshold, further enhancing tolerance toward unreliable observations in high wind speed regimes.

\subsubsection{TC boundary layer physical constraints}
\label{sec:physics}

Reconstructing wind direction from pure scalar wind speed is the core challenge QiFeng faces. Even though the diffusion prior provides the statistical structure of TC wind fields, the nonlinear likelihood constraint alone still cannot fully resolve the directional ambiguity of $(u,v)$: for each observed wind speed value, infinitely many direction combinations satisfy the constraint with equal probability. Therefore, we designed three differentiable TC boundary layer physical regularizations, encoding known low-level dynamical features of TCs. These are calculated as gradients with respect to $\hat{\mathbf{x}}_0$ at each denoising step and injected into the reverse diffusion process to resolve the directional ambiguity at a physical level. All constraints operate on a 4$\times$ downsampled grid ($64\times 64$, 6\,km resolution), which both reduces computational overhead and matches the characteristic scales of TC boundary layer processes.

\paragraph{(i) Low-level divergence constraint.}
The magnitude of horizontal divergence in the TC boundary layer $O(10^{-5}\,\mathrm{s}^{-1})$ is much smaller than the vorticity magnitude $O(10^{-3}\,\mathrm{s}^{-1})$ \cite{holton2013dynamics, kepert2001tcbl}. This constraint restricts the combinational patterns of spatial gradients in $u$ and $v$---local convergence or divergence cannot be excessively strong---thus indirectly constraining the spatial change rate of wind directions to prevent physically unreasonable wind field structures (e.g., abrupt divergence zones or frontal-like discontinuities):
\begin{equation}
	\mathcal{L}_\text{div} = \frac{1}{N}\sum_{p,q}\left(\frac{\partial u}{\partial x}\bigg|_{p,q}+\frac{\partial v}{\partial y}\bigg|_{p,q}\right)^{\!2},
	\label{eq:div-loss}
\end{equation}
where partial derivatives are calculated using central differences. Note that the grid's $i$-axis (row index) increases southward, hence $\partial v/\partial y = -\partial v/\partial i$. Significant boundary layer convergence (magnitudes up to $O(10^{-3}\,\mathrm{s}^{-1})$) indeed exists in the TC eyewall region, but this constraint operates on a 4$\times$ downsampled grid (6\,km resolution). At this scale, the area-averaged divergence is far smaller than local values at the convective scale; the 6\,km grid effectively filters out convective-scale convergence/divergence signals, retaining the divergence features of the mesoscale circulation, which is indeed much smaller than vorticity in mature TC vortices \cite{shapiro1983asymmetric, kepert2001tcbl}. Furthermore, $\mathcal{L}_\text{div}$ is a quadratic penalty rather than a hard constraint ($\text{div}=0$), physically permitting the existence of necessary non-zero divergence, and only suppressing extreme divergence/convergence patterns that are incompatible with TC dynamics.

\paragraph{(ii) Bounded inflow angle constraint.}
Based on over 1600 dropsonde observations, \citet{zhang2012inflow} established comprehensive climatological statistics for TC 10\,m inflow angles: the full-sample mean is $22.6^\circ\pm 2.2^\circ$, with a systematic radial dependence. The inner-core region ($r<2\,R_\text{max}$) typically shows $\sim$10$^\circ$--15$^\circ$, while the outer region ($r>3\,R_\text{max}$) rises to $\sim$25$^\circ$--30$^\circ$. Although individual observations have a wider scatter (in extreme cases deviating to below 5$^\circ$ or above 40$^\circ$, for instance, the inflow angle significantly increases on the down-shear side in strong shear environments \cite{zhang2013asymmetric}), in a climatological sense, TC boundary layer inflow angles are robustly constrained within a finite interval. The inflow angle $\alpha$ directly dictates the allocation ratio between the tangential wind $V_t$ and radial wind $V_r$, meaning that constraining the inflow angle is equivalent to constraining the radial-tangential decomposition of wind direction relative to the TC center---which is precisely the key to resolving directional ambiguity. Decomposing the wind into the radial component $V_r$ (pointing toward the TC center is positive, i.e., convergence) and the tangential component $V_t$ (cyclonic tangential is positive), the inflow angle is defined as $\alpha=\arctan(-V_r/V_t)$. Based on these climatological statistics, we impose soft boundary penalties (rather than hard truncation), allowing the model to freely deviate from typical ranges when local physical processes necessitate it:
\begin{equation}
	\mathcal{L}_\text{inflow} = \frac{1}{|\mathcal{M}|}\sum_{(p,q)\in\mathcal{M}}\left[\frac{\text{ReLU}(\alpha_\text{min}-\alpha)}{\pi/4}\right]^2+\left[\frac{\text{ReLU}(\alpha-\alpha_\text{max})}{\pi/4}\right]^2,
	\label{eq:inflow-loss}
\end{equation}
where $\alpha_\text{min}=10^\circ$, $\alpha_\text{max}=35^\circ$, and the normalization factor $\pi/4$ ensures moderate gradient magnitudes. The quadratic structure of this loss function means that the gradient contribution of inflow angles falling within $[\alpha_\text{min},\,\alpha_\text{max}]$ is strictly zero, free of any bias; inflow angles deviating from this interval are subjected to a progressively increasing quadratic penalty. The larger the deviation, the stronger the penalty, but it remains a finite value. This contrasts fundamentally with fixed parametric models (e.g., the Holland inflow angle formula $\alpha(r)=\alpha_\text{max}(r/R_\text{max})\exp(1-r/R_\text{max})$), which enforces a single unique inflow angle value at every grid point. Instead, $\mathcal{L}_\text{inflow}$ provides only a loose physical envelope, allowing the specific radial distribution and azimuthal variation to be determined entirely by the diffusion prior and observation likelihood. The chosen boundary values have been moderately expanded relative to the climatological statistical ranges in \cite{zhang2012inflow} (inner-core mean $\sim$12$^\circ$, outer mean $\sim$27$^\circ$) to accommodate individual differences among storms and variability across atypical azimuthal sectors.

The mask $\mathcal{M}$ excludes regions where the inflow angle definition is unstable or its physical behavior deviates from typical statistics in two ways: (a) Weak wind regions ($|\mathbf{V}|<15$\,m\,s$^{-1}$), where the wind field is dominated by large-scale environmental flows and convective randomness. Here, the inflow angle definition lacks clear physical meaning and is numerically unstable ($\alpha\to\pm 90^\circ$ as $V_t\to 0$); (b) Near the TC center ($r<18$\,km, i.e., 3 downsampled pixels), because the directional organization of winds inside the eye is weak, the definition of the radial direction degenerates at the center, and inflow angle observation statistics inside the eyewall are sparse. Additional numerical stability designs include: detaching the radial/tangential unit vectors from the computational graph to prevent the $1/r$ factor from producing gradient explosions at the TC center; and using a $\text{atan2}$ implementation with a numerical $\epsilon$ instead of naive $\arctan$ to ensure finite gradients as $V_t\to 0$.

\paragraph{(iii) Positive cyclonic vorticity constraint.}
TC circulation at 10\,m altitude is predominantly characterized by strong cyclonic vorticity (Northern Hemisphere $\zeta>0$), a fundamental dynamic property of TCs as warm-core vortices. This constraint ensures the physical correctness of the reconstructed wind field's rotation direction on a global scale: it imposes a one-sided penalty only on anticyclonic vorticity ($\zeta<0$) while placing no restrictions on the magnitude of positive vorticity. This ensures the correct rotational direction while granting the diffusion prior full freedom to express cyclonic circulations of varying intensities and spatial distributions:
\begin{equation}
	\mathcal{L}_\text{vort} = \frac{1}{|\mathcal{M}'|}\sum_{(p,q)\in\mathcal{M}'}\min\!\left(\left[\frac{\text{ReLU}(-\zeta_{p,q})}{10^{-3}}\right]^2,\,10\right),
	\label{eq:vort-loss}
\end{equation}
where $\zeta=\partial v/\partial x - \partial u/\partial y$ is the relative vorticity, and the normalization factor $10^{-3}\,\mathrm{s}^{-1}$ corresponds to typical vorticity magnitudes in the TC inner core. The truncation upper bound of 10 prevents extreme gradient values during high-noise diffusion steps (large $\sigma_t$), ensuring numerical stability during optimization. The mask $\mathcal{M}'$ excludes regions where wind speed is $<$10\,m\,s$^{-1}$. This design considers two aspects: in weak wind regions, vorticity is numerically dominated by finite difference noise, carrying limited physical meaning; additionally, localized anticyclonic vorticity (such as residual signals of anticyclonic outflow layers) can indeed exist in the outer environmental flow fields of TCs. Imposing constraints here is both unnecessary and likely to introduce unreasonable biases. For Southern Hemisphere TCs (cyclonic = $\zeta<0$), the observation grid coordinates are flipped along the north-south axis before QiFeng sampling, keeping the model consistently operating under the Northern Hemisphere convention (cyclonic = $\zeta>0$), and the reconstructed wind field is flipped back after output---avoiding the need to design separate physical constraints for the two hemispheres.

\paragraph{Complementarity and design principles of the three constraints.}
The total physical loss is $\mathcal{L}_\text{phys}=\mathcal{L}_\text{div}+\mathcal{L}_\text{inflow}+\mathcal{L}_\text{vort}$. The three constraints provide directional guidance from different dimensions: the divergence constraint acts on the spatial gradient structures of $u$ and $v$ in Cartesian coordinates, suppressing unreasonable divergence/convergence patterns; the inflow angle constraint directly dictates the radial-tangential allocation ratio in a polar perspective, which is the core mechanism resolving scalar-vector directional ambiguity; and the vorticity constraint ensures the physical correctness of the rotation direction on a global scale. The three are applied in different coordinate systems and spatial scales, with mutually orthogonal mathematical forms---divergence and vorticity constrain $\partial u/\partial x + \partial v/\partial y$ and $\partial v/\partial x - \partial u/\partial y$, respectively, constituting the trace and antisymmetric components of the velocity gradient tensor; the inflow angle constrains the deflection angle of the wind vector relative to the radial direction in the natural coordinate system. This complementarity ensures that even in the complete absence of directional observations, the directional degrees of freedom of $(u,v)$ are effectively constrained.

In design, all three constraints adhere to the ``loose envelope'' principle: each constraint encodes the most fundamental and robust physical features in TC boundary layer dynamics (quasi-non-divergent, bounded inflow angle, cyclonic rotation), rather than attempting to precisely specify the exact structure of the wind field. All constraints utilize a soft penalty format (quadratic or one-sided ReLU), where gradient contributions within the constraint interval are zero and increase asymptotically outside it, allowing the diffusion prior and observation likelihood to maintain dominance within physically reasonable bounds. The three constraints are merged with equal weights, because each loss has already been calibrated to its respective physical magnitude normalization factor (divergence: implicit in quadratic magnitude; inflow angle: $\pi/4$; vorticity: $10^{-3}\,\mathrm{s}^{-1}$). This keeps their gradient norms on the same order of magnitude under typical TC conditions. The strength of the total physical loss is controlled by a single hyperparameter $\lambda$ (Section~\ref{sec:lambda-selection}), avoiding the difficulties of jointly tuning multiple hyperparameters.

\subsubsection{Guided reverse sampling}
\label{sec:guided-sampling}

Based on the EDM's Heun ODE solver \cite{karras2022edm}, QiFeng injects the observation likelihood and physical constraints described above as gradient guidance at each time step. The total likelihood score (Eq.~\eqref{eq:score-like}):
\begin{equation}
	\mathbf{s}_\text{like} = -\nabla_{\mathbf{x}_t}\!\left(\mathcal{L}_\text{obs} + \lambda\,\mathcal{L}_\text{phys}\right)
	\label{eq:score-like}
\end{equation}
is obtained through automatic differentiation ($\lambda=15$, see Section~\ref{sec:lambda-selection}). Each step first executes $C=2$ rounds of Langevin corrector steps (step size $\delta=(\tau\sigma_t)^2$, $\tau=0.3$), alternately fusing the prior score $\mathbf{s}_\text{prior}=(\hat{\mathbf{x}}_0-\mathbf{x})/\sigma_t^2$ and likelihood score $\mathbf{s}_\text{like}$ at the current noise level, gradually drawing the sample toward the conditional posterior distribution. This is then advanced to the next noise level by the guided Heun predictor. In the Heun predictor, the likelihood score directly modifies the ODE's drift term $\mathbf{d}_i = (\mathbf{x}-\hat{\mathbf{x}}_0)/\sigma_i - \sigma_i\cdot\mathbf{s}_\text{like}$, ensuring that observational and physical constraints remain continuously effective across noise-level state transitions. The full sampling process employs an $N=64$ step Karras noise schedule ($\sigma_{\max}=80$, $\sigma_{\min}=0.002$, $\rho=7$), with the baseline observation noise $\sigma_y$ set to 0.01 in OSSEs and 0.03 for real data (reflecting the larger uncertainty in true observations), and the approximation error weight $\gamma=0.001$. The complete workflow is shown in Algorithm~\ref{alg:qifeng}.

\begin{algorithm}[htbp]
	\caption{QiFeng Inference: Physics-Guided Nonlinear Likelihood Assimilation}
	\label{alg:qifeng}
	\KwInput{EDM network $D_{\boldsymbol\theta}$; CYGNSS observations $\{(i_k,j_k,y_k,\sigma_k)\}_{k=1}^K$; hemisphere flag}
	\KwHyper{$N$, $\sigma_\text{min}$, $\sigma_\text{max}$, $\rho$, $\sigma_y$, $\gamma$, $C$, $\tau$, $\lambda$}
	\KwOutput{Reconstructed TC vector wind field $\hat{\mathbf{x}}\in\mathbb{R}^{2\times 256\times 256}$}
	\BlankLine
	$\sigma_i \leftarrow \bigl(\sigma_\text{max}^{1/\rho}+\frac{i}{N-1}(\sigma_\text{min}^{1/\rho}-\sigma_\text{max}^{1/\rho})\bigr)^{\rho}$, $i=0,\dots,N$ \tcp*{Karras schedule}
	\lIf{Southern Hemisphere}{$i_k \leftarrow 255-i_k$\tcp*{Flip to NH convention}}
	$\mathbf{x} \leftarrow \sigma_0\cdot\boldsymbol\epsilon$, $\boldsymbol\epsilon\sim\mathcal{N}(\mathbf{0},\mathbf{I})$\;
	\BlankLine
	\For{$i=0$ \KwTo $N-1$}{
		$\delta \leftarrow (\tau\cdot\sigma_i)^2$\;
		\For(\tcp*[f]{Langevin corrector}){$c=1$ \KwTo $C$}{
			$\hat{\mathbf{x}}_0 \leftarrow D_{\boldsymbol\theta}(\mathbf{x};\,\sigma_i)$\;
			$\mathbf{s}_\text{prior} \leftarrow (\hat{\mathbf{x}}_0 - \mathbf{x})/\sigma_i^2$\;
			Compute $V_{\text{eff},k}$ (Eq.~\eqref{eq:veff}), $\mathcal{L}_\text{obs}$ (Eq.~\eqref{eq:obs-loss}, Huber robust loss)\;
			$\mathcal{L}_\text{phys} \leftarrow \mathcal{L}_\text{div}+\mathcal{L}_\text{inflow}+\mathcal{L}_\text{vort}$ \tcp*{Eqs.~\eqref{eq:div-loss}, \eqref{eq:inflow-loss}, \eqref{eq:vort-loss}}
			$\mathbf{s}_\text{like} \leftarrow -\nabla_{\mathbf{x}}(\mathcal{L}_\text{obs}+\lambda\,\mathcal{L}_\text{phys})$\;
			$\mathbf{x} \leftarrow \mathbf{x}+\frac{\delta}{2}(\mathbf{s}_\text{prior}+\mathbf{s}_\text{like})+\sqrt{\delta}\,\boldsymbol\xi$, $\boldsymbol\xi\sim\mathcal{N}(\mathbf{0},\mathbf{I})$\;
		}
		\tcp{Guided Heun predictor}
		$\hat{\mathbf{x}}_0,\,\mathbf{s}_\text{like} \leftarrow \textsc{ScoreGuide}(\mathbf{x},\,\sigma_i)$\;
		$\mathbf{d}_i \leftarrow (\mathbf{x}-\hat{\mathbf{x}}_0)/\sigma_i - \sigma_i\cdot\mathbf{s}_\text{like}$\;
		$\tilde{\mathbf{x}} \leftarrow \mathbf{x}+(\sigma_{i+1}-\sigma_i)\,\mathbf{d}_i$\;
		\If{$i<N-1$}{
			$\hat{\mathbf{x}}_0',\,\mathbf{s}_\text{like}' \leftarrow \textsc{ScoreGuide}(\tilde{\mathbf{x}},\,\sigma_{i+1})$\;
			$\mathbf{d}_{i+1} \leftarrow (\tilde{\mathbf{x}}-\hat{\mathbf{x}}_0')/\sigma_{i+1} - \sigma_{i+1}\cdot\mathbf{s}_\text{like}'$\;
			$\mathbf{x} \leftarrow \mathbf{x}+(\sigma_{i+1}-\sigma_i)\cdot\frac{1}{2}(\mathbf{d}_i+\mathbf{d}_{i+1})$\;
		}
		\lElse{$\mathbf{x} \leftarrow \tilde{\mathbf{x}}$}
	}
	\lIf{Southern Hemisphere}{Flip $\mathbf{x}$ along $i$-axis}
	\Return $S\cdot\mathbf{x}$ \tcp*{Denormalize to m\,s$^{-1}$}
\end{algorithm}

\subsection{Observation Error Model}
\label{sec:obs-error}

CYGNSS YSLF retrievals are known to exhibit positive biases and sensitivity saturation at high wind speeds \cite{morris2017cygnss, ruf2019cygnss}. To enable QiFeng to adaptively handle observations of varying quality, we designed a wind-speed-dependent observation-wise error model tailored to CYGNSS GNSS-R characteristics. The error standard deviation for each observation point $k$ contains two components (Eq.~\eqref{eq:sigma-model}):
\begin{equation}
	\sigma_k = \sigma_\text{base}(y_k) + 0.5\,|\Delta t_k|,
	\label{eq:sigma-model}
\end{equation}
where $\Delta t_k$ is the time offset (in hours) between the observation time and the target analysis time, and the temporal decay coefficient of 0.5\,m\,s$^{-1}$\,h$^{-1}$ reflects the hourly scale evolution of TC structures. The baseline error $\sigma_\text{base}$ is a piecewise linear function (Eq.~\eqref{eq:sigma-base}):
\begin{equation}
	\sigma_\text{base}(y) = \begin{cases}
		2.0 & y \leq 30\,\text{m\,s}^{-1}, \\
		2.0 + 0.3(y-30) & 30 < y \leq 50\,\text{m\,s}^{-1}, \\
		8.0 + 0.5(y-50) & y > 50\,\text{m\,s}^{-1}.
	\end{cases}
	\label{eq:sigma-base}
\end{equation}
The physical basis for this three-segment design is: (a) for $y\leq 30$\,m\,s$^{-1}$, CYGNSS YSLF retrieval accuracy is high, with a baseline error of 2.0\,m\,s$^{-1}$, consistent with values reported in the literature \cite{ruf2019cygnss}; (b) in the 30--50\,m\,s$^{-1}$ range, L-band GNSS-R sea surface scattering cross-sections begin to saturate, retrieval sensitivity drops \cite{ruf2024characterization}, and the error increases linearly (slope 0.3), reaching $\sim$8\,m\,s$^{-1}$ at 50\,m\,s$^{-1}$; (c) for $y>50$\,m\,s$^{-1}$, the signal fully enters the saturation zone, severe precipitation attenuation exacerbates, and the error grows with a steeper slope (0.5). This model embeds likelihood guidance via the $(\sigma_k/S)^2$ term in Eq.~\eqref{eq:veff}: for example, an observation point at 60\,m\,s$^{-1}$ has $\sigma_k\approx 13$\,m\,s$^{-1}$, making its $V_{\text{eff}}$ approximately 40 times larger than that of a 20\,m\,s$^{-1}$ observation point ($\sigma_k=2$\,m\,s$^{-1}$), thus significantly diminishing its gradient contribution---preventing unreliable extreme wind speed observations from dominating the reconstruction results.

In the OSSE mode (where simulated observations are derived from HWRF ground truths without retrieval biases), the slope above 30\,m\,s$^{-1}$ is reduced to 0.15, reflecting the more ideal observational conditions in the OSSE environment.

\subsection{Data}
\label{sec:data}

\paragraph{HWRF Simulation Data.}
Training data for the diffusion model were derived from innermost-nest simulations of the HWRF \cite{tallapragada2016hwrf} model at a spatial resolution of 1.5\,km. Because the official HWRF servers only retain downloads for the most recent two days, making historical data difficult to access directly, we established automated scripts to continuously collect and organize HWRF analysis fields over 2023--2024; the resulting dataset covers 117 TCs in total. The $(u,v)$ 10\,m wind components were extracted from the 1013\,hPa pressure level of each simulation's f000 initial analysis field (to avoid systematic forecast biases), cropped to a $256\times 256$ grid (384\,km $\times$ 384\,km) centered on the TC origin, normalized to $[-1,1]$ (by dividing by $S=150$\,m\,s$^{-1}$), and stored as $(2,256,256)$ tensors, yielding 1{,}130 HWRF initial field snapshots in total. To ensure strict data independence and prevent leakage, the partitioning strategy was based on TC names---all analysis snapshots of the same TC belonged exclusively to either the training set or the test set. The training set contains 89 TCs ($\sim$764 snapshots), and the test set contains 28 TCs (366 snapshots).

\paragraph{NICAM Pre-training Data.}
The pre-training stage utilized Non-hydrostatic Icosahedral Atmospheric Model (NICAM) \cite{satoh2008nicam, satoh2014nicam} global cloud-resolving simulations. NICAM is a global non-hydrostatic atmospheric model based on an icosahedral grid, with a horizontal resolution of $\sim$14\,km, capable of directly resolving deep convective processes without convective parameterization, thereby reproducing TC mesoscale structures (eyewalls, spiral rainbands, etc.). We used the 30-year (1979--2009) NICAM climate simulation long-sequence dataset organized by \citet{matsuoka2023tcdata}, which tracked 2,463 TC paths and provided over 55,000 instantaneous TC snapshots covering all major basins and intensity ranges. Although NICAM's spatial resolution ($\sim$14\,km) is lower than the target HWRF analysis fields (1.5\,km), its sample size is vastly larger ($\sim$12 times), providing ample diversity for the diffusion model to learn universal statistical structures of TC circulations (asymmetries, spiral rainbands, multi-scale vortices). In preprocessing, we extracted near-surface 10\,m wind fields $(u_{10}, v_{10})$ from the NICAM outputs, projected them to the same azimuthal equidistant projection coordinate system as HWRF using the best-track-provided TC center positions as references, cropped horizontal subdomains of 384\,km $\times$ 384\,km consistent with HWRF, and interpolated them to a $256\times 256$ grid (equivalent resolution 1.5\,km), ensuring spatial coverage and grid specifications were identical to the fine-tuning stage. The variable normalization strategy remained exactly the same as for the HWRF dataset (dividing by $S=150$\,m\,s$^{-1}$).

\paragraph{CYGNSS Observation Data.}
QiFeng utilizes the Young Seas with Limited Fetch (YSLF) wind speed products from the CYGNSS L2 v3.2 Climate Data Record as observational inputs. Compared to Fully Developed Seas products, YSLF is specifically optimized for sea states dominated by young wind waves in TC environments, offering higher retrieval accuracy and a wider dynamic range under high wind speed conditions \cite{ruf2019cygnss, ruf2024characterization}.

Observation collection is anchored to the TC center positions and times reported by IBTrACS \cite{knapp2010ibtracs}, gathering all CYGNSS specular points falling within the 384\,km $\times$ 384\,km domain within a $\pm$3\,h time window. The $\pm$3\,h window balances spatial coverage (OCS pass rate 24.9\%, vs.\ 6.7\% for $\pm$1\,h) against temporal smoothing (TC intensity can change 10--15\,kt over 6\,h), consistent with CYGNSS's median revisit ($\sim$3\,h) and the SCG product design \cite{ruf2019cygnss, mayers2023scg, demaria2005ships}. A quantitative sensitivity analysis across four time windows ($\pm$1--6\,h) is provided in Supplementary Table~S3.

Quality control is split into two levels: (1) Observation-wise quality control---based on the \texttt{yslf\_sample\_flags} variable in CYGNSS L2 V3.2 products, all observations marked with \texttt{fatal\_*} are removed (a bitmask is automatically constructed according to the \texttt{flag\_masks} and \texttt{flag\_meanings} attributes in the CF-1.6 metadata \cite{cygnss2023atbd}), and a physical upper limit of 70\,m\,s$^{-1}$ is imposed to remove spurious extremes (this threshold refers to the nominal dynamic range upper limit of CYGNSS L2 YSLF retrievals \cite{cygnss2023atbd}); (2) Inter-track consistency quality control---observations within the $\pm$3\,h window are clustered into different satellite overpass tracks by time gaps, the average wind speed of each track is calculated, and outlier tracks whose average wind speed deviates by more than 20\,m\,s$^{-1}$ from the track median are removed. This eliminates contradictory constraints between tracks caused by rain contamination or retrieval anomalies. Processed observations are mapped to the $256\times 256$ grid coordinates.

During spatiotemporal matching, each specular point's corresponding TC center position is linearly interpolated from the IBTrACS path based on its observation time. The observation point is then corrected to TC-relative coordinates, thereby eliminating the influence of TC translational motion on observation positioning within the $\pm$3\,h window, employing a methodology similar to that of SCG products \cite{mayers2023scg}.

\subsection{Evaluation Metrics}
\label{sec:metrics}

To systematically evaluate QiFeng's reconstruction quality, this study employs the following evaluation metric system.

\paragraph{TC Structural Parameters.}
Five key TC structural parameters are extracted from the reconstructed $(u,v)$ wind fields: $V_{\max}$, RMW, and wind radii R34/R50/R64. Each wind radius is calculated as the average of its four quadrants (NE/NW/SW/SE), consistent with the definition of IBTrACS USA wind radii. RMW is measured from the TC eye center (the local wind speed minimum point within the domain). To eliminate the confounding influence of varying eye locations among data sources, all wind fields have their eyes aligned to the grid center before calculating structural parameters. Reconstructed values are compared with IBTrACS best-track values or HWRF ground truths (in OSSEs), reporting MAE (mean absolute error), Bias (Estimate $-$ Reference), and RMSE (root mean square error).

\paragraph{Wind Direction Fidelity.}
Since QiFeng's core innovation lies in reconstructing full vector fields from scalar observations, wind direction accuracy is a critical evaluation dimension. The wind is decomposed into a tangential component $V_t$ (cyclonic tangential is positive) and a radial component $V_r$ (pointing toward the TC center is positive), and the RMSE of azimuthally averaged tangential/radial wind profiles ($V_t$-RMSE, $V_r$-RMSE) are calculated respectively. Wind direction errors employ circular statistics to correctly handle $0^\circ$/$360^\circ$ wraparound, reporting Dir-RMSE, Dir-MAE, and Dir-Bias. The average difference of inflow angle $\alpha=\arctan(-V_r/V_t)$ in the 30--150\,km radial range ($\Delta\alpha_{\mathrm{inflow}}$) is used to evaluate the accuracy of boundary layer inflow structures.

\paragraph{Point-by-point Matching Validation.}
Comparisons with independent observations such as dropsondes and TDR employ a spatiotemporal matching method: within a given time window (dropsonde $\pm$0.5\,h, TDR $\pm$1.5\,h) and spatial window, QiFeng reconstructed values are matched point-by-point with observation values, reporting MAE, Bias, RMSE, and Pearson correlation $R$ for $U$/$V$ components and scalar wind speed.

\subsection{TDR Altitude Correction Method}
\label{sec:tdr-correction}

TDR 0.5\,km wind speeds were corrected to 10\,m using a radially dependent reduction factor based on the mean boundary layer profiles of \citet{franklin2003dropsonde}: a larger factor ($f \approx 0.76$) near the eyewall ($r/R_\mathrm{MW} \le 1.5$), where the super-gradient boundary layer jet is strongest \cite{kepert2001tcbl, kepert2006bljet}, and a smaller one ($f \approx 0.87$) in the outer vortex ($r/R_\mathrm{MW} \ge 3.0$), with linear transition in between. Similar flight-level-to-surface reduction methods have been operationally adopted by NHC/HRD \cite{powell2009surface, uhlhorn2007sfmr}. Full details and the correction equation are provided in Supplementary Section~S9.

\subsection{OSSE Experimental Design}
\label{sec:osse-design}

We designed OSSEs using 366 HWRF test-set TC snapshots as ground truth, simulating CYGNSS observations by extracting scalar wind speeds along predefined tracks with additive Gaussian noise. Three orbital coverage scenarios were designed (Supplementary Figure~S6): Scenario A (full inner-core, $\sim$357 points), Scenario B (no-eye, $\sim$283 points, the most representative of real CYGNSS coverage), and Scenario C (eyewall-only, $\sim$148 points). All scenarios use an along-track spacing of $\sim$6\,km and $\lambda=15$ (Section~\ref{sec:lambda-selection}).


\section*{Data availability}
The CYGNSS Level-2 Climate Data Record (v3.2) wind speed data are available from NASA PO.DAAC (\url{https://podaac.jpl.nasa.gov/CYGNSS}). IBTrACS best-track data are available from NOAA National Centers for Environmental Information (\url{https://www.ncei.noaa.gov/products/international-best-track-archive}). ERA5 reanalysis data are available from the Copernicus Climate Data Store (\url{https://cds.climate.copernicus.eu}). CCMP v3.1 ocean surface wind data are available from Remote Sensing Systems (\url{https://www.remss.com/measurements/ccmp/}). The NICAM TC simulation dataset is available via the reference \cite{matsuoka2023tcdata}. NOAA HRD Tail Doppler Radar composite wind fields and dropsonde data are available from the Hurricane Research Division (\url{https://www.aoml.noaa.gov/hrd/}). SAR-derived ocean surface wind fields are available from the CYCLOBS dataset (\url{https://cyclobs.ifremer.fr/}). The HWRF analysis fields were collected from the NOAA Environmental Modeling Center operational servers.

\section*{Code availability}
The custom code for the QiFeng framework, including the diffusion model training, physics-guided nonlinear likelihood assimilation, and evaluation scripts, is available from the corresponding authors upon reasonable request.

\bibliographystyle{unsrtnat}
\bibliography{references}

\section*{Acknowledgements}
This work was supported by the National Natural Science Foundation of China (42530402, 42192561), the Special Project--Original Exploration (Grant 42450254), the NSFC--FDCT Grants 62361166662, the National Key R\&D Program of China (2023YFC3503400, 2022YFC3400400), the Project of State Key Laboratory of Satellite Ocean Environment Dynamics (SOEDZZ2527), and Innovation Group Project of Engineering Guangdong Laboratory (Zhuhai) (311024004, SML2023SP202).

\section*{Author contributions}
X.Han conceived and designed the study, developed the methodology, implemented all code, conducted all experiments, performed data curation and quality control for all datasets, carried out all analyses and validations, created all figures and tables, and wrote the manuscript. X.Li supervised the research, coordinated multi-source observational data, contributed to data interpretation, and revised the manuscript. J.Yang supervised the overall research direction, provided GNSS-R remote sensing expertise, secured the major funding, administered the project, and revised the manuscript. Z.Niu assisted with TC meteorological data processing and IBTrACS matching. G.Han contributed to the interpretation of TC--ocean interaction results and revised the manuscript. J.Wang contributed to the deep learning model architecture design and provided computational resources. W.Huang assisted with typhoon observational data curation and meteorological analysis. Y.Zheng contributed to the CYGNSS observation quality control and data preprocessing. H.Ni assisted with CYGNSS data collection and specular point gridding. Y.Wang assisted with model validation experiments and evaluation metric calculation. W.Tao contributed to the computational framework optimization. L.Aouf contributed expertise on satellite ocean surface wind remote sensing methodology and revised the manuscript. S.Peng provided high-performance computing resources and revised the manuscript. D.Chen contributed to the physical oceanographic interpretation, provided guidance on TC boundary layer dynamics, and revised the manuscript.

\section*{Competing interests}
The authors declare no competing interests.

\end{document}